\documentclass{aa}  

\usepackage{graphicx}
\usepackage{txfonts}

\usepackage{tcolorbox}

\usepackage{float}
\usepackage{soul}

\usepackage{caption}
\usepackage{subcaption}

\usepackage{hyperref}
\usepackage{xcolor}

\definecolor{mycitecolor}{RGB}{0, 0, 255} 
\hypersetup{
    colorlinks=true,
    linkcolor=mycitecolor, 
    citecolor=mycitecolor, 
    urlcolor=black, 
}

\begin{document} 

\title{A Synthetic Population of Ultra-Luminous X-ray Sources}
\subtitle{Optical--X-ray Correlation}

   \author{Lutendo Nyadzani
          \inst{1},
          Soebur Razzaque\inst{1,2,3}, and Justin D.\ Finke\inst{4}}
          
   \institute{
   Centre for Astro-Particle Physics (CAPP) and Department of Physics, University of Johannesburg, PO Box 524, Auckland Park 2006, South Africa\\
   \email{lnyadzani@uj.ac.za};
    \email{srazzaque@uj.ac.za}
    \and
    Department of Physics, The George Washington University, Washington, DC 20052, USA
    \and
    National Institute for Theoretical and Computational Sciences (NITheCS), Private Bag X1, Matieland, South Africa
    \and
    U.S. Naval Research Laboratory, Code 7653, 4555 Overlook Ave. SW, Washington, DC 20375-5352, USA \\
    \email{justin.d.finke.civ@us.navy.mil}
    }

   \date{Received}
 
  \abstract
 {This paper presents an analysis of the 
   predicted
   optical-to-X-ray spectral index ($\alpha_{\rm ox}$) within the context of ultra-luminous X-ray sources (ULXs) associated with stellar mass black holes and neutron stars. We use the population synthesis code \textsc{cosmic} to simulate the evolution of binary systems and investigate the relationship between UV and X-ray emission during the ULX phase, namely the $\alpha_{\rm ox}$ relation. 
   The study investigates the impact of metallicity on $\alpha_{\rm ox}$ values. Notably, it predicts a significant anti-correlation between $\alpha_{\rm ox}$ and UV luminosity ($L_{\rm UV}$), consistent with observations, with the slope of this relationship varying with metallicity for BH-ULXs. The NS-ULX population shows a relatively consistent slope around $-0.33$ across metallicities,
with minor variations. The number of ULXs decreases with increasing metallicity, consistent with observational data, and the X-ray luminosity function shows a slight variation in its slope with metallicity, exhibiting a relative excess of high-luminosity ULXs at lower metallicities. Inclusion of beaming effect in the analysis shows a significant impact on the XLF and $\alpha_{\rm ox}$, particularly at high accretion rates, where the emission is focused into narrower cones.
Furthermore, the study finds that UV emission in ULXs is predominantly disk-dominated, which is the likely origin of the $\alpha_{\rm ox}$ relation, with the percentage of disk-dominated ULXs increasing as metallicity rises.  

}
   \maketitle
%

\section{Introduction}

Ultra-luminous X-ray sources (ULXs) are extragalactic point-like sources that exhibit extraordinarily high luminosities in the X-ray $(L_{\rm x}>10^{39}$ erg/s), exceeding the Eddington luminosity for black holes (BHs) of mass $\approx 10M_\odot$. The first ULXs were discovered using the Einstein Observatory, designed to obtain images of distant galaxies {\citep[for a review, see][]{fabbiano89}.} Early X-ray surveys revealed a considerable quantity of these sources. \citet{fabbiano89} identified 16 sources with X-ray luminosities $L_{\rm x}>10^{39}$ erg/s. To date, over 1800 ULXs have been identified, increasing our knowledge and understanding of these systems \citep{kaaret2017ultraluminous, walton2022multimission, KING2023101672}. 

Since their discovery, the debate on what powers these systems has been a continuous topic of both theory and observation. Theoretically, there are three possibilities as to how these systems reach such high luminosities; 1) Accretion by intermediate mass black holes (IMBH), $10^2$~M$_{\odot}$ to $10^5~$M$_{\odot}$, \citep{colbert1999nature}: this increases the mass of the BH and the system accretes at sub-Eddington rate; 2) Beamed emission from stellar mass BH or NS \citep{georganopoulos2002external}: this avoids Super-Eddington accretion and keeps the mass of the compact object below 100 M$_{\odot}$; 3) Super-Eddington accretion \citep[e.g.,][]{begelman01,finke07,skadowski2015powerful}: accretion onto stellar mass BHs or NSs emitting with a super-Eddington luminosity. Note that a system can simultaneously be beamed and supercritical. Observations have shown evidence for IMBH \citep{lasota2011origin,mezcua2013radio}, stellar mass BH \citep{middleton2013bright}, and pulsations detected in a few systems indicate the presence of a NS \citep{bachetti2014ultraluminous,furst2016discovery} as the primary object in these systems. It has been suggested that ULXs could be the progenitors of compact object mergers detected by LIGO \citep{inoue16,finke17,mondal20}.

The debate on ULXs centres around the mass of the compact objects and their accretion rate. While X-ray data has been instrumental in our understanding of ULXs, it has limitations. It often does not provide information about the companion of the compact object. Identifying the companion can offer valuable insights into the mass of these binary systems. An example of this approach's significance can be found in the confirmation of the existence of Cygnus X-1, the first known Galactic BH \citep{murdin1971optical}. ULXs also emit radiation in other wavelengths, including the ultraviolet (UV) and optical bands \citep{gladstone2013optical}.

Previous observational studies have provided valuable insights into the connection between X-ray and UV radiation in ULXs, suggesting a common origin or interconnected emission mechanisms. \citet{sonbas2019evidence} investigated the relationship between the X-ray and ultraviolet wavebands in several ULXs with known optical counterparts by comparing the ratio of flux densities and by utilising the spectral optical–X-ray index, $\alpha_{\rm ox}$. Their analysis reveals a significant anti-correlation between $\alpha_{\rm ox}$ and the UV luminosity at $2500$ Å  ($L_{\rm 2500 Å}$). However, due to the complexity of ULX systems and the limitations of observational data, it is challenging to fully explore and understand the correlation between X-ray and UV radiation based solely on observations. Therefore, simulated studies that follow the evolution of binary systems and model the emission processes to provide detailed insights into the spectral behaviour are essential. Simulated studies enable us to disentangle the complex interplay of physical mechanisms in astrophysical sources, allowing us to explore a wide parameter space and investigate various emission scenarios. By employing a population synthesis code, we simulate a range of ULX scenarios and systematically study the spectral correlation between X-ray and UV radiation. 

In this paper, we use the population synthesis code \textsc{cosmic} \citep{breivik2020cosmic} to simulate the evolution of binary systems and study the relationship between the UV and X-ray emission of ULXs. We explore effects of metallicity on the synthetic population of ULXs and the $\alpha_{\rm ox}$ relation for BH-ULXs and NS-ULXs, and compare with observational data. The layout of the paper is as follows: in Section \ref{synthetic_section}, we discuss our \textsc{cosmic} simulation set-up; the UV and X-ray emission models are discussed in Section \ref{emission_section}; and in Section \ref{results_section} we present and discuss our results. We summarise and conclude our study in Section \ref{summary_section}.

\section{Synthetic binary population}
\label{synthetic_section}

To study the correlation between X-ray and UV radiation in ULXs we employed the \textsc{cosmic} \citep{breivik2020cosmic} binary population synthesis code. \textsc{cosmic} combines theoretical models of stellar evolution, binary interactions, and compact object mergers to simulate the formation and evolution of binary systems. We model the evolution of binaries starting from zero-age main sequence (ZAMS) until the formation of the X-ray binary systems.  

In our simulations, we drew the primary (most massive) ZAMS mass, $M_1$, from a Kroupa initial mass function \citep[IMF;][]{kroupa1993distribution}.  This IMF is given by $dN/dM_1\propto M_1^{-\alpha}$, with $\alpha=1.3$ for $0.08~{\rm M}_{\odot} < M_1\leq 0.5~ {\rm M}_{\odot}$; $\alpha=2.2$ for $0.5 ~{\rm M}_{\odot} < M_1 \leq 1~ {\rm M}_{\odot}$; and $\alpha=2.7$ for $M_1 > 1.0~{\rm M}_{\odot}$.  The ZAMS mass of the secondary star, $M_2$ (ranging from $0.5~{\rm M}_{\odot}$ to $150~{\rm M}_{\odot}$), was drawn from a flat distribution of the binary mass ratio $q = M_2/M_1$ within the range of [0.1, 1.0]. The orbital period ($P$) was drawn from a distribution $f(\log_{10} P/d)\approx (\log_{10} P/d)^{-0.55}$, with $\log_{10} P/d$ ranging from 0.15 to 5.5, and the eccentricity $(e)$ was drawn from a distribution that followed $f(e)\approx e^{-0.42}$ within the interval of [0.0, 0.9] \citep{sana2013vlt}.  

In our study, we simulated $2\times10^6$ binary systems with varying metallicities, ranging from $0.5\%$ of the solar metallicity ($Z_{\odot}$) to $1.5~Z_{\odot}$. The precise value of $Z_{\odot}$ remains uncertain \citep{vagnozzi2017solar}, we adopt  $Z_{\odot} = 0.02$ as the benchmark value for consistency with previous studies \citep[e.g.,][]{mondal20}. Additionally, we set the binary fraction to be $100\%$ for primary mass greater than 10 M$_{\odot}$ \citep{sana2013vlt}. The conﬁgurations in our standard \textsc{cosmic} model parameters are as follows: 
\begin{enumerate}
\item Common Envelope (CE) phase: CE efﬁciency of $\alpha_{\rm CE} = 1$, we follow the pessimistic scenario by letting stars without a core-envelope boundary to automatically lead to merger during the CE phase \citep{belczynski2007rarity}.

\item Supernova model: To estimate the mass of the final compact object after the supernova explosion, we follow the rapid supernova model {\citep{fryer2012compact}.} 

\item Accretion: Studies using simulations have demonstrated that mass accretion onto a compact object can exceed the Eddington limit {\citep{mckinney2014three}.} It is now considered possible for super-Eddington accretion to occur in X-ray binaries with both BH and NS as compact objects \citep{van2019chandra, vasilopoulos20202019}. Therefore, we allow for super-Eddington accretion to reach values of up to 1000 times the Eddington limit.

\item Critical mass ratios $(q_c)$: $q_c$, is the mass ratio between the donor and accretor at the beginning of Roche lobe overflow (RLOF). It is used to determine whether a system will experience a CE phase or continue with stable mass transfer (MT) based on the mass ratio of the components at the beginning of RLOF. Mass transfer through RLOF is allowed for thermal time-scale only if the mass ratio at the beginning of RLOF is below the $q_c$. If the mass ratio $\geq q_c$, mass transfer occurs on a dynamical time-scale, resulting in the initiation of the CE phase. To determine $q_c$, we follow the treatment in Section 5.1 of \citet{belczynski2008compact}. 

\end{enumerate}

We sample a population of main-sequence binary stars that are evolved from ZAMS until the formation of X-ray binary systems. Our ULX population is selected based on the value of X-ray luminosity, $L_{\rm x}>10^{39}$ erg/s. In the next section, we discuss how our X-ray and UV luminosities are calculated.

\section{Emission from accretion disk and donor star}
\label{emission_section}

\subsection{Total X-ray emission}
The X-ray emission from ULXs primarily originates from the accretion processes onto the compact objects, either stellar-mass BHs or NSs. In these systems, a massive donor star transfers mass to the compact object through RLOF or wind accretion. The high luminosities observed in ULXs, exceeding the Eddington limit for typical stellar-mass BHs, suggest the presence of highly efficient accretion processes. Accretion models play a crucial role in understanding the X-ray emission from ULXs and require careful consideration of physical parameters and radiative processes.

In a compact binary system, mass is transferred from one star to another. This process leads to the creation of an accretion disk. We calculate an isotropic  bolometric X-ray luminosity ($L_{\rm X, iso})$ from accretion as \citep{shakura1973black,poutanen2007supercritically}
\begin{equation}
    L_{\rm X,iso} = \begin{cases}
        L_{\rm Edd}[1+\ln\dot{m}] & \text{if } \dot{m}>1  \\
        \eta \dot{M}c^2 & \text{if } \dot{m} \leq 1 \,
    \end{cases}\ .
    \label{eq:lxiso}
\end{equation}
Here $\dot{M}$ is the mass accretion rate, $c$ is the speed of light, $\dot{m}$ is the ratio $\dot{M}/\dot{M}_{\rm Edd}$ between the mass accretion rate, and  the Eddington accretion rate $\dot M_{\rm Edd}=12L_{\rm Edd}/c^2
$, 
$L_{\rm Edd}$ is the Eddington luminosity of a star or compact object with mass $M$ given by
\begin{equation}
    L_{\rm Edd} = 1.26\times 10^{38}\left(\frac{M}{\rm M_{\odot}}\right)~\text{erg s}^{-1}\,,
    \label{eq:ledd}
\end{equation}
$\eta$ is the accretion efficiency, defined as
\begin{equation}
    \eta_{\rm NS}=\frac{GM_{\rm NS}}{c^2R_{\rm NS}} \,,
\end{equation}
for NSs, and
\begin{equation}
    \eta_{\rm BH}= 1- E(R_{\rm ISO}) \,,
\end{equation}
for BHs, where $G$ is Newton's gravitational constant. The function $E(R)$ is given by
\begin{equation}
    E(R) = \frac{R^2-2\frac{GM}{c^2}R+a\frac{GM}{c^2}\left(\frac{GM}{c^2}R\right)^{1/2}}{R\left[R^2-3\frac{GM}{c^2}R+2a\frac{GM}{c^2}\left(\frac{GM}{c^2}R\right)^{1/2}\right]^{1/2}} \,,
\end{equation}
where $a$ is the BH spin and $R_{\rm ISO}$ is the radius of the innermost stable orbit
\begin{equation}
R_{\text{ISO}} = \frac{R_{\text{Sch}}}{2} \left( 3 + Z_2 - \left[ (3 - Z_1) (3 + Z_1 + 2Z_2) \right]^{1/2} \right)
\end{equation}

\begin{equation}
Z_1 = 1 + \left( 1 - \frac{4a^2}{R_{\text{Sch}}^2} \right)^{1/3} \left[ \left( 1 + \frac{2a}{R_{\text{Sch}}} \right)^{1/3} + \left( 1 - \frac{2a}{R_{\text{Sch}}} \right)^{1/3} \right]
\end{equation}

\begin{equation}
Z_2 = \left( \frac{3a^2}{R_{\text{Sch}}^2} + Z_1^2 \right)^{1/2}
\end{equation}
where $R_{\rm Sch}=2GM/c^2$ is the Schwarzschild radius \citep{brown2000theory}. 

 When the accretion rate is high, the luminosity can be focused into narrow cones, resulting in greater observed luminosity compared to the isotropic emission. This phenomenon is referred to as ``beaming''. In order to explore the potential impact of the beaming factor, we follow \citet{king2009masses} to calculate the beaming factor ($b$), which is given by
\begin{equation}
    b= \begin{cases}
        \frac{73}{\dot{m}^2}& \text{if } \dot{m}>8.7\\
        1 & \text{if } \dot{m}<8.7 \,.
    \end{cases}\
    \label{eq:beam}
\end{equation}
The beamed X-ray luminosity is then given by:
\begin{equation}
    L_{\rm X, b} = \frac{L_{\rm X, iso}}{b}\,,
\end{equation}
where $L_{\rm X,iso}$ is the X-ray luminosity in Equation~(\ref{eq:lxiso}).

\subsection{ULX spectrum}

 In addition to their intense X-ray emission, ULXs emit in optical/UV. The presence of UV emission in conjunction with X-ray radiation opens a window to understanding the broader radiative processes at play in these enigmatic systems. This dual emission may hold clues to the intricate interconnections between the various components of the accretion process, the accretion disk, companion stars, and the surrounding environment. To study ULXs in both the UV and X-ray bands, it is essential to consider contributions from both the accretion disk and the stellar companion. The X-ray emission primarily originates from the accretion disk, whereas the UV emission has two components: one from the disk and another from the stellar companion. In this section, we describe the models utilised in our calculations. 

\subsubsection{Accretion disk }
\label{disksection}

 ULX spectral modelling has been a focus in several studies using models such as \textsc{diskbb} \citep{1984PASJ...36..741M} and \textsc{diskir} \citep{2008MNRAS.388..753G,2009MNRAS.392.1106G}. These models often imply the presence of high-mass black holes in the ULX systems, which are not the focus of this study. \citet{vinokurov2013ultra} proposed a physically motivated model by examining the spectral energy distribution (SED) of supercritical accretion disks (SCADs) and applied this model to fit observed SEDs of five ULXs.

According to the SCAD model, the accretion disks become thick and develop a strong wind, forming a wind funnel above the disk at distances less than the spherization radius, with a temperature dependence given by \( T \propto r^{-1/2} \). The disk's luminosity in this region is approximated by \( L_{\text{bol}} \approx L_{\text{Edd}} \ln (\dot{m}) \). Outside the spherization radius, the disk remains thin, and its total luminosity equals the Eddington luminosity, \( L_{\text{Edd}} \). The thin disk heats the wind from below, and from the inner side of the funnel, the wind is heated by the supercritical disk \citep[see Fig.\ 2 of][]{vinokurov2013ultra}.

When applied to ULX spectra, the SCAD model suggests stellar-mass BHs as the compact object, whereas the \textsc{diskir} model often yields much higher BH masses, suggesting IMBHs. The SCAD model accounts for the high-energy curvature observed in ULX spectra and the presence of outflows without invoking IMBHs. It explains the observed luminosities and spectral features using stellar-mass BHs undergoing supercritical accretion, consistent with the lack of a large population of IMBHs observationally. Since all our simulated ULXs have either stellar-mass BHs or NSs as the compact objects, we use the SCAD model to describe the supercritical accretion disks in our simulated ULX population. The temperature of the disk in this model is given by 

\begin{equation}
T(r) = T_{\rm in }
\begin{cases}
    (r \sin \theta_{f})^{-1/2}, & 1 \leq r \leq r_{\rm sp} \\
     \Big[\frac{f_{\rm out}}{\sin \theta_{f}} (1 + \ln \dot{m})\Big]^{1/4}r^{-1/2}, & r_{\rm sp} \leq r \leq r_{\rm ph}
\end{cases}
\label{Vsp}
\end{equation}
where $T_{\rm in}^4 = {L_{\rm Edd}\sin(\theta _f)}/({4\pi\sigma R_{\rm in}^2})$, $R_{\rm in}= 3R_{\rm Sch}=6 GM/c^2$,  $M$ is the compact object mass, $\theta_f $ is the funnel angle, $\sigma$ is the Stefan-Boltzmann constant, $f_{\rm out}$ is the fraction of bolometric flux thermalised in the disk, and $r=R/R_{\rm in}$ is the radius of the accretion disk. Here, \( r_{\rm sp} = {\dot{M} }/\dot{M}_{\rm Edd} \) is the spherization radius, and \( r_{\rm ph} = {\dot{M} \kappa}/({4 \pi v_w \cos \theta_f R_{\rm in}}) \) is the photospheric radius,  where \(\kappa\) is the opacity (\(\kappa\approx \kappa_{\rm es}= 0.34~ \text{cm}^2/\text{g}\) for fully ionised gas) and \(v_w = 1000\) km s$^{-1}$ is the wind velocity. With our \textsc{cosmic} data, this leaves only $f_{\rm out}$ and $\theta_{f}$ as free parameters in this model. The effect of these two free parameters is discussed in the results section. The luminosity, at a given frequency $\nu$ 
 and radius \( r \) within a thickness of \( dr \) of the disk can be expressed as:

\begin{equation}
dL_\nu(r) = 2\pi R^2_{\rm in} \cdot r B_\nu( T(r)) \cdot dr
\end{equation}
where 
\begin{equation}
\label{planckeqn}
B_\nu( T) = \frac{2h\nu^3}{c^2}\cdot \frac{1}{e^{\frac{h\nu}{kT}} - 1}
\end{equation}
is the Planck function describing blackbody radiation at temperature \( T \) and a given frequency $\nu$. Here $h$ is Planck's constant and $k$ is Boltzmann's constant.  Therefore, the total luminosity as a function of frequency, considering the temperature gradient, is:
\begin{equation}  
L_\nu =  2 \pi R^2_{\rm in} \int_{1}^{r_{\rm ph}} B_\nu(T(r)) \, r \, dr\ .
\label{Lv}
\end{equation}

\subsubsection{Stellar companion }

 Another component of the ULX spectrum is the companion star, which mainly contributes to the UV band. Observations have indicated that the most likely donor stars are massive and hot OB-type stars, which can contribute significantly to the UV emission  \citep{liu2004optical,roberts2008new,gladstone2013optical}. Other studies have found evolved red super-giant companions \citep{heida2015discovery,heida2019discovery}. This emission can also be attributed to the high-energy radiation absorbed and re-emitted by the donor star's surface \citep{middleton2015spectral}. The UV luminosity from a companion star of radius ($R_{\star}$) and effective temperature ($T_{\star}$) is obtained by,
\begin{equation}
    L_{\rm UV,\star} = \pi R^2_{\star} \nu  B_\nu (T_{\star}) 
    \label{eq:uv_spectrum}
\end{equation}
where we use $\lambda=c/\nu=2500~\text{\AA}$. 
The UV contribution from the disk ($L_{\rm UV, disk}$) is obtained from Equation (\ref{Lv}) at  $\lambda=2500~\AA$. Then, the total UV emission from the ULX is given by
\begin{equation}   
L_{\rm UV} = L_{\rm UV,\star} + L_{\rm UV, disk}\ .
\end{equation}

\section{Results and discussion }
\label{results_section}
 Our study aims to simulate a population of ULXs, investigate their properties, and compare them to the observed population. In this section we show our results and compare them to the observational data where available. 

\subsection{Number of ULXs}

 Various studies have explored the relationship between the number of ULXs, star formation rate (SFR), and metallicity. Using the population synthesis code StarTrack, \citet{wiktorowicz2017origin} simulated a robust correlation between the number of ULXs and the SFR history. Meanwhile, \citet{mondal20}, also using StarTrack, found in their simulations an increase in the number of ULXs with decreasing metallicity, specifically in the context of stars born at the same time (i.e., a single burst). In an X-ray survey of a sample of 66 galaxies, \citet{mapelli2011remnants} found that there is no significant direct correlation between $N_{\rm ULX}$ and metallicity, however, when normalising the number of ULXs by the SFR, ($N_{\rm ULX}$/SFR), a marginally significant anti-correlation with metallicity is observed. This indicates that lower metallicity environments may produce more ULXs per unit of star formation compared to higher metallicity environments. The anti-correlation suggests that metallicity has some influence on the efficiency of ULX formation, but this effect is secondary to the influence of SFR. \citet{mapelli2011remnants} also found the power-law index of $\alpha=0.55^{+0.21}_{-0.19}$ for $N_{\rm ULX}$/SFR where the number of ULX is normalised by the star formation rate and $\alpha=0.16^{+0.28}_{-0.28}$ for $N_{\rm ULX}$ as a function of metallicity.

\begin{figure}[h]
    \centering
    \includegraphics[scale=0.35]{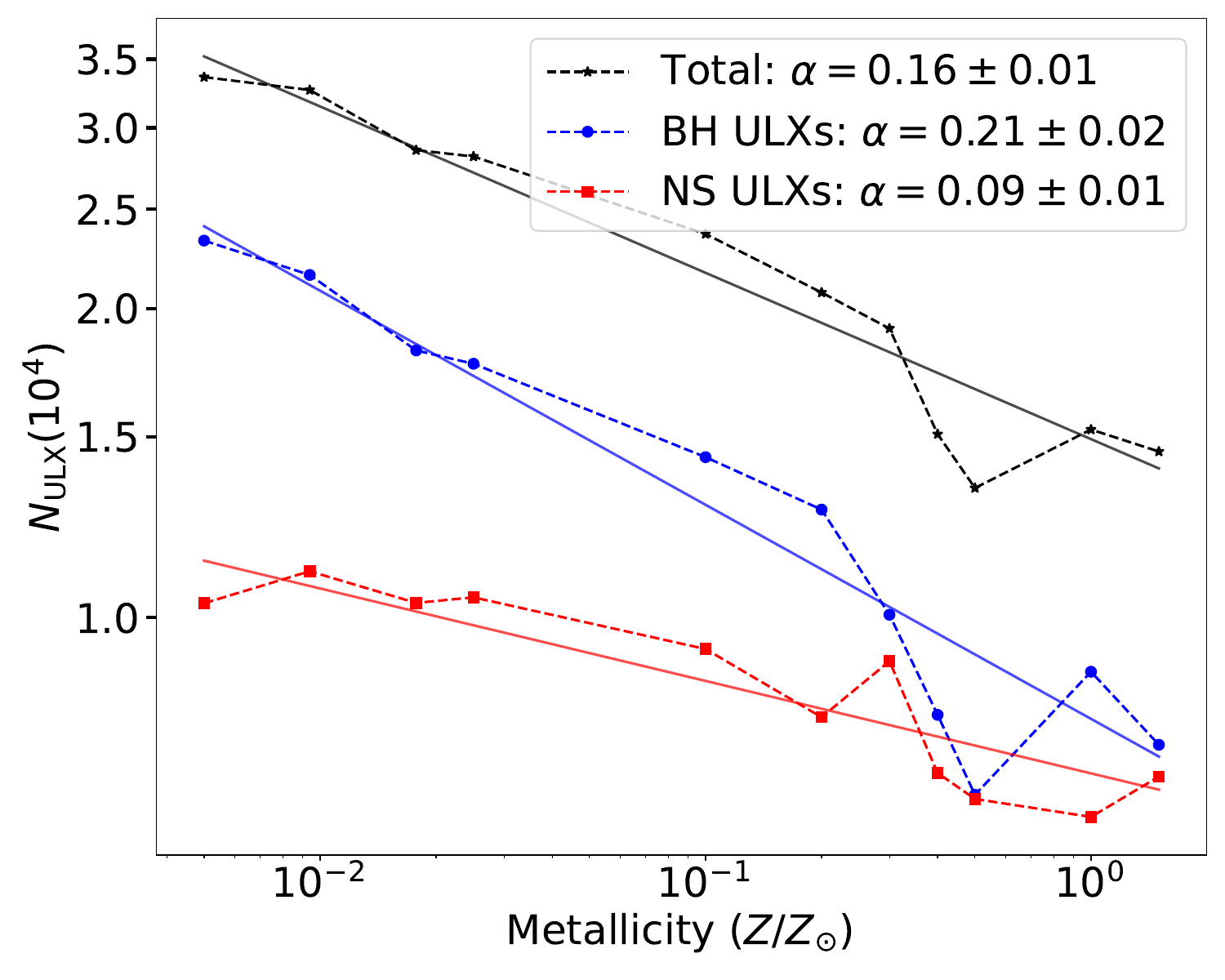}
    \caption{The number of ULXs formed in our simulation ($N_{\rm ULX}$) as a function of metallicity. The y-axis indicates the number of ULXs derived from the simulations, with emphasis placed on the trend's shape rather than the absolute values. The solid lines represent fits to the simulated data, with the corresponding slopes mentioned in the legend.}
    \label{fig:Number}
\end{figure}

In Fig.~\ref{fig:Number}, we show $N_{\rm ULX}$ as a function of metallicity for our simulations. Note that we define $N_{\rm ULX}$ from our synthetic population at the end of the ULX phase. Studying the time evolution of ULX population properties will be the focus of a future study. The total number of ULXs, as well as the sub-populations of BH-ULXs and NS-ULXs, show a declining trend with increasing metallicity. The slope for BH-ULXs is steeper ($\alpha = 0.21 \pm 0.02$) compared to the total ULX population ($\alpha = 0.16 \pm 0.01$) and NS-ULXs ($\alpha = 0.09 \pm 0.01$). At higher metallicities, the populations of NS-ULXs and BH-ULXs become comparable. The steeper decline for BH-ULXs indicates that the formation of BH-ULXs is more sensitive to metallicity compared to NS-ULXs. The shallower slope for NS-ULXs suggests that while metallicity does affect their formation, it is not as critical as for BH-ULXs. This aligns with the theoretical understanding that massive metal-poor stars are more likely to form massive BHs through direct collapse \citep{heger2003massive}. Our power-law index of $\alpha=0.16\pm 0.01$ for combined population of BH- and NS-ULXs is consistent with the slope from \citet{mapelli2011remnants}.


The parameters $f_{\rm out}$ and $\theta_f$ in Equation~(\ref{Vsp}) play a crucial role in shaping the UV and X-ray emission characteristics of ULXs. A higher fraction of the bolometric flux thermalised in the disk (i.e., larger $f_{\rm out}$) results in stronger UV emission. Conversely, for any observer viewing the system down the funnel formed by the disk/wind combination, smaller $\theta_f$ angles will result in enhanced X-ray emission by directing more of the X-ray emission into the increasingly narrow funnel. 
For our simulated populations, we set $f_{\rm out}$ to the average value derived from the systems analysed in \citet{vinokurov2013ultra}, adopting $f_{\rm out} = 0.03$. A detailed discussion of the impact of varying $f_{\rm out}$ on our results can be found in Section~\ref{freeP}. This value of $f_{\rm out}$ ensures that a small but significant fraction of the bolometric flux is thermalised in the accretion disk, contributing to UV emission. Given the high accretion rates typical in ULXs, this thermalization fraction effectively reproduces the observed emission without overestimating the flux.

We study our population with $\theta_f$ set to $45^{\circ}$ and treat this as non-beamed population (see Section~\ref{Nobeam}). It should be noted that \(\theta_f = 45^\circ\) was selected as a standard, intermediate value within the allowed range of $0^\circ-90^\circ$ for exploratory purposes and does not imply any inherent physical property of the system. In Section~\ref{beamsec}, we calculate $\theta_f$ from the accretion rate using the beaming model by \citet{king2009masses} and compare the two scenarios.

\subsection{Results with constant $\theta_f$}\label{Nobeam}

\subsubsection{X-ray luminosity function}

\begin{figure}[H]
\centering
\includegraphics[width=1.\linewidth]{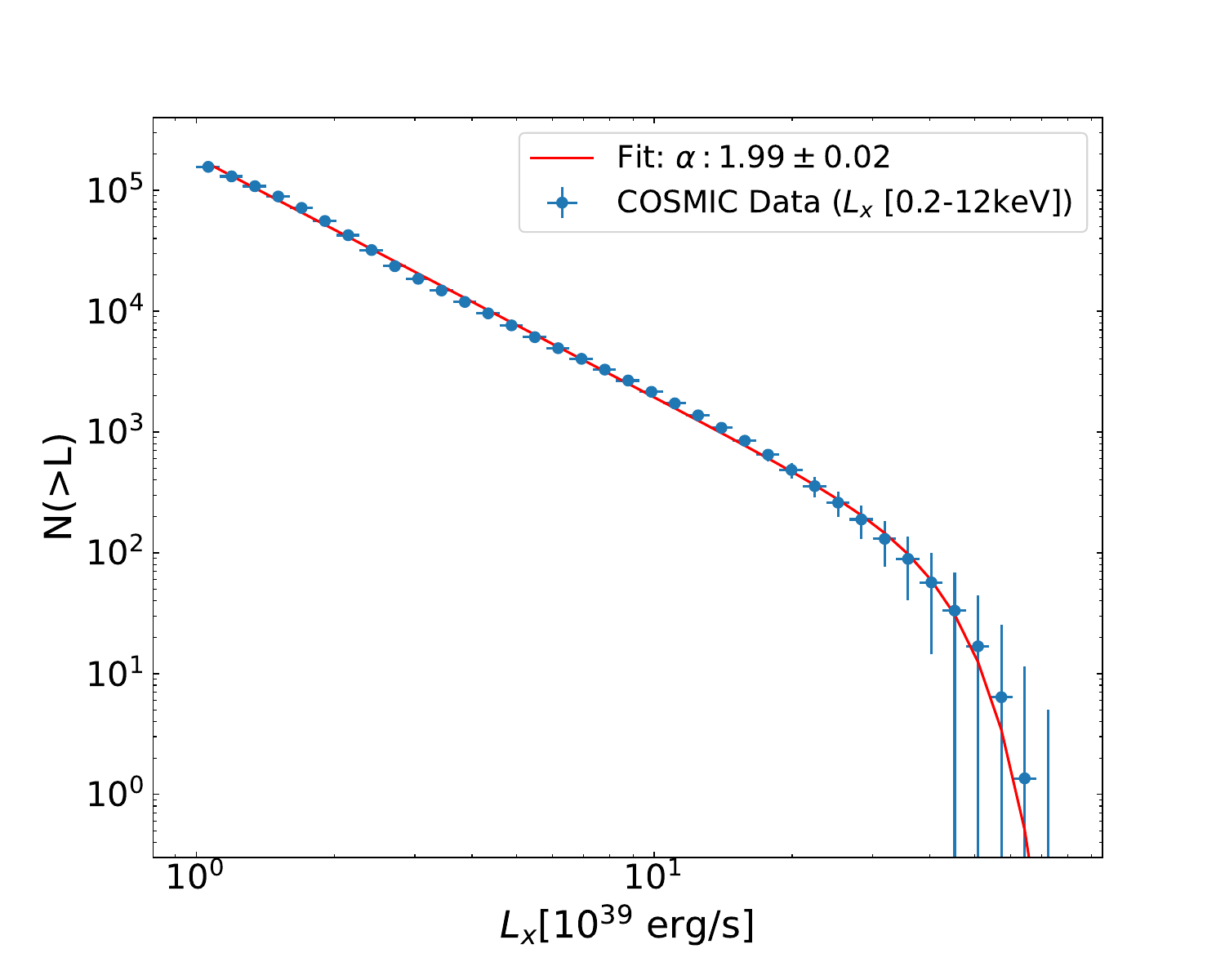}
\caption{Cumulative XLF of the simulated ULX population at 40\% $Z_{\odot}$ in the 0.2-12 keV band.
}
\label{fig:XLF2}
\end{figure}

One approach to studying the ULX population involves studying the X-ray luminosity function (XLF) for these sources. Fig.~\ref{fig:XLF2} shows the cumulative XLF of the simulated ULX population at 40\% of $Z_{\odot}$ in the 0.2-12 keV band. We fit this population with a function
\begin{equation}
N(>L_X) = CL_{\rm X}^{-\alpha}\exp{[-(L_{\rm X}/L_s)^{\beta}]} \,,
\label{eq:XLF}
\end{equation}
and obtained power-law index 
 \( \alpha = 1.99 \pm 0.01 \), $\beta = 3.92\pm 3.70$, and $L_s = (43.7\pm 9.94)\times10^{39}$~erg~s$^{-1}$  at this metallicity. However, the power-law index changes with metallicity. The XLF for a high-metallicity population is  steeper than a low-metallicity population; it ranges from $\alpha=1.77\pm 0.01$ at $0.05Z_{\odot}$ to $2.15\pm 0.38$ at $1.5Z_{\odot}$, showing a relative excess of high-luminosity ULXs at lower metallicities (see Table \ref{tab:fit_XLF}). Similarly, \citet{lehmer2021metallicity} observed an increasing number of luminous sources with declining metallicity, indicating that low-metallicity environments have more high-luminosity high-mass X-ray binary (HMXB) sources.

\begin{table}[h!]
\centering
\caption{XLF fit parameters. }
\begin{tabular}{|c|c|c|c|}
\hline
    
    Metallicity& slope (\(\alpha \))& \(L_S\)        & \(\beta\)       \\
    (\% Z$_\odot$)&&($\times 10^{39}$erg/s)&\\
    \hline
   0.5 &1.77$\pm$ 0.01 & 59.5 \(\pm\) 7.37      & 3.66 \(\pm\) 1.81      \\
   
    2.5&1.83$\pm$ 0.02 & 58.5 \(\pm\) 9.98       & 3.88 \(\pm\) 2.82       \\
    10& 1.87$\pm$ 0.03& 62.2 \(\pm\) 14.2      & 3.63 \(\pm\) 3.06       \\
    20&1.93$\pm$ 0.02 & 63.6 \(\pm\) 17.4       & 3.86 \(\pm\) 4.14       \\
    30&1.96$\pm$ 0.02 & 50.1 \(\pm\) 11.5      & 3.60 \(\pm\) 3.08       \\
   40 & 1.99$\pm$ 0.02& 43.7 \(\pm\) 9.94      & 3.92 \(\pm\) 3.70       \\
   
    50&1.95$\pm$ 0.32 & 20.8 \(\pm\) 8.41       & 1.32 \(\pm\) 0.84       \\
    
   100 &1.87$\pm$ 0.80 & 3.49 \(\pm\) 4.31      & 1.58 \(\pm\) 0.43        \\
   
    150&2.15$\pm$ 0.38 & 2.83 \(\pm\) 1.32       & 0.58 \(\pm\) 0.43       \\
    \hline
\end{tabular}
\tablefoot {Values of XLF parameters \(\alpha\), \(L_s\) and \(\beta\) with their respective uncertainties for different metallicities, assuming an observer is always looking down the funnel. We used a fixed value of the accretion disk funnel angle $\theta_f = 45^\circ$.}
\label{tab:fit_XLF}
\end{table}

\subsubsection{Optical--X-ray spectral index ($\alpha_{\rm ox}$)}

The X-ray-to-UV ratio is described as the interband spectral index \citep{tananbaum1979x}
\begin{equation}
    \alpha_{\rm ox}=\frac{\log_{10}{(L_{\rm \nu, x}/L_{\rm \nu,UV})}}{\log_{10}{(\nu_{\rm x}/\nu_{\rm UV})}} \,,
    \label{eq:alphaOX}
\end{equation}
where both \(L_{\nu, UV}\) and \(L_{\nu, X}\) are expressed as \textit{monochromatic luminosities} (in units of erg/s/Hz) defined at $\nu_{\rm UV}=1.2 \times 10^{15}$ Hz ($2500~\text{\AA}$) and $\nu_{\rm x}=4.8 \times 10^{17}$ Hz (2~keV) respectively, following the work of \citet{sonbas2019evidence}, 
 who found an anti-correlation between $\alpha_{\rm ox}$ and $L_{\nu,\rm UV}$ in observational data. In the context of the SCAD model, discussed in Sec.~\ref{disksection}, such a relationship is expected, where the UV luminosity is dominated by emission from the funnel and depends on the accretion rate ($\dot m$), while the X-ray luminosity is dominated by emission from the inner accretion disk and does not depend on $\dot m$ (see figure~3 in \citet{vinokurov2013ultra}).  Both the components depend on the mass of the primary in the same way.  Systems with greater $\dot m$ have greater UV luminosity, but the same X-ray luminosity (for a given primary mass $M$).  So as $\dot m$ increases, $L_{\nu,\rm UV}$ also increases, and the ratio $L_{\nu,X}/L_{\nu,\rm UV}\propto \alpha_{\rm ox}$ decreases.  This leads to the anti-correlation between $\alpha_{\rm ox}$ and $L_{\nu,\rm UV}$. The diversity of this relation, however, depends on various factors such as the metallicity of the systems, accretion disk's wind velocity, fraction of flux thermalisation in the disk, etc. We present those results here.

\begin{figure}[ht]
    \centering
    \includegraphics[scale=0.37]{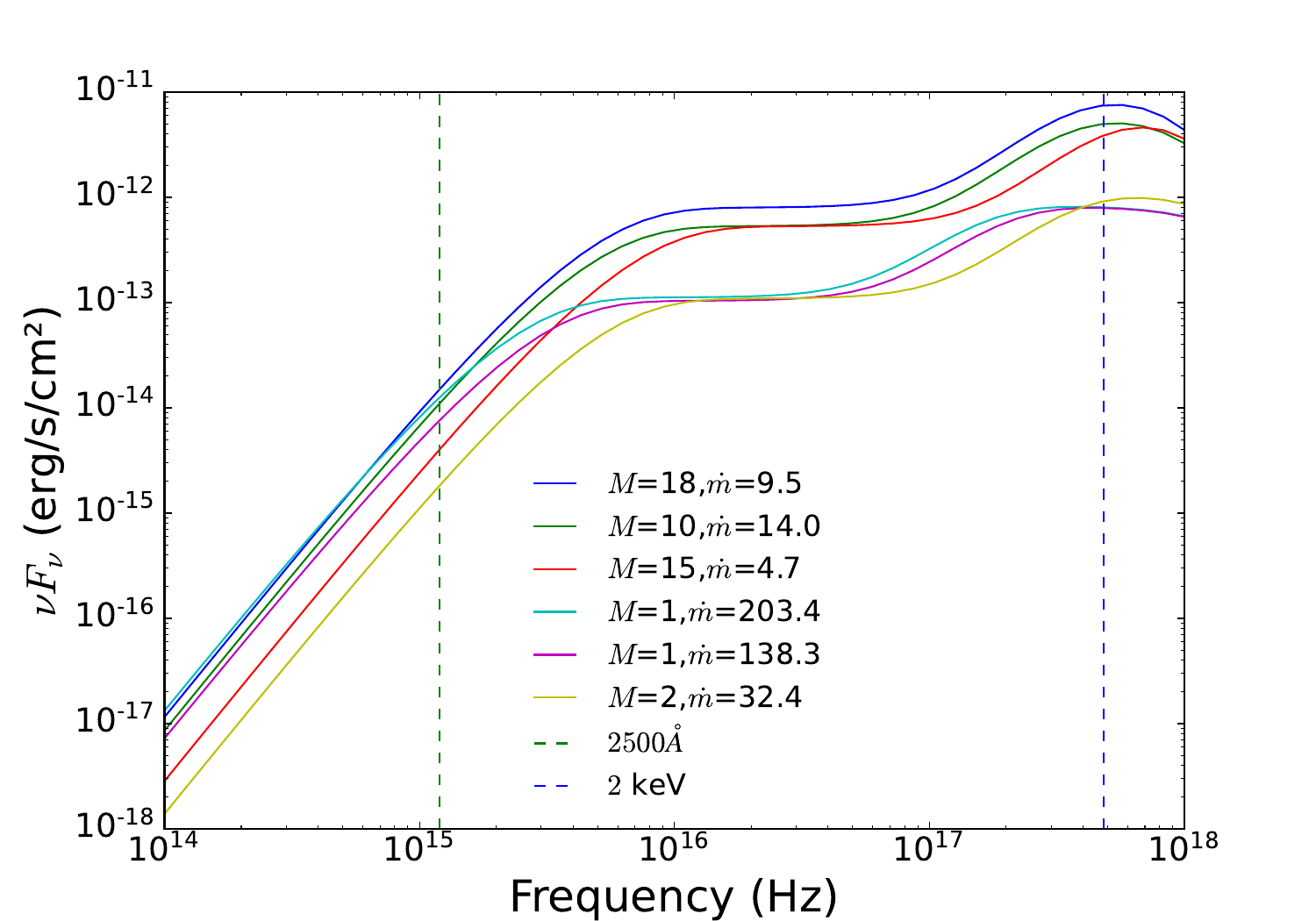}
    \caption{Spectrum of 3 BH-ULX and 3 NS-ULX systems from our simulation, showing the mass of the compact object ($M$) and the ratio $\dot{m}=\dot{M}/\dot{M}_{\rm Edd}$. Flux is calculated for all systems located at 3 Mpc, and for all curves $f_{\rm out}=0.03$ and $\theta_f = 45^{\circ}$.}
    \label{fig:spectrum}
\end{figure}

Fig.~\ref{fig:spectrum} presents the assumed SEDs for three BH-ULXs and three NS-ULXs generated from
our simulations \citep[cf.\ Fig.~3 of][]{vinokurov2013ultra}. The spectra are calculated at a 
distance of 3 Mpc using Equation (\ref{Lv}) without the contribution from the stellar companion. The figure also includes the key parameters for each system: $M$ and $\dot m$. We have shown in Fig.~\ref{fig:SED_mdot} the effect of varying $\dot m$ for a constant $M$, where it is evident that the 2~keV X-ray flux is independent of $\dot m$, a crucial point made above for the $\alpha_{\rm ox}$ and $L_{\nu,\rm UV}$ anti-correlation. Other studies have shown that the ULX wind velocity can be as high as $0.1c-0.3 c$ \citep{pinto2023ultra, pinto2020xmm, 10.1093/mnras/sty1626}. Increasing the wind velocity from 1000 km~s$^{-1}$ to 60,000 km~s$^{-1}$ ($0.2c$) leads to a reduction in flux at lower frequencies and the disappearance of the UV shoulder. However, the X-ray emission at higher frequencies is unaffected by the change in wind velocity (see Fig.~\ref{fig:sed_02c}). This results in more positive $\alpha_{\rm ox}$ values and steeper $\alpha_{\rm ox}-L_{\nu,\rm UV}$ relation (see Fig.~\ref{fig:alpha_02c}).

\begin{figure}
    \centering
    \begin{subfigure}{0.5\textwidth}
        \centering
        \includegraphics[scale=0.3]{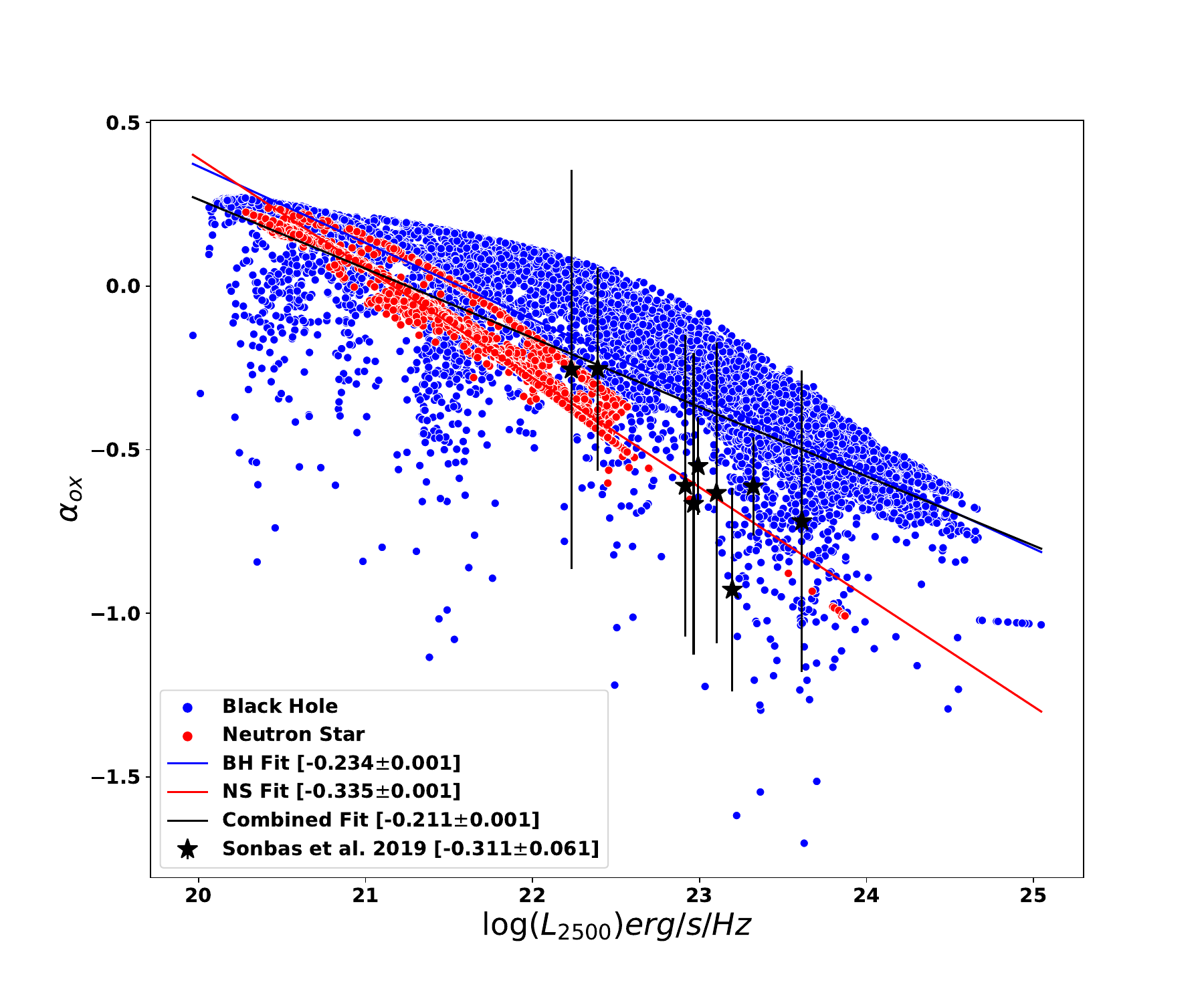}
        \vspace{-0.6cm}
        \label{fig:subfig-a}
    \end{subfigure}
    \hfill
    \begin{subfigure}{0.5\textwidth}
        \centering
        \vspace{-0.6cm}
        \includegraphics[scale=0.3]{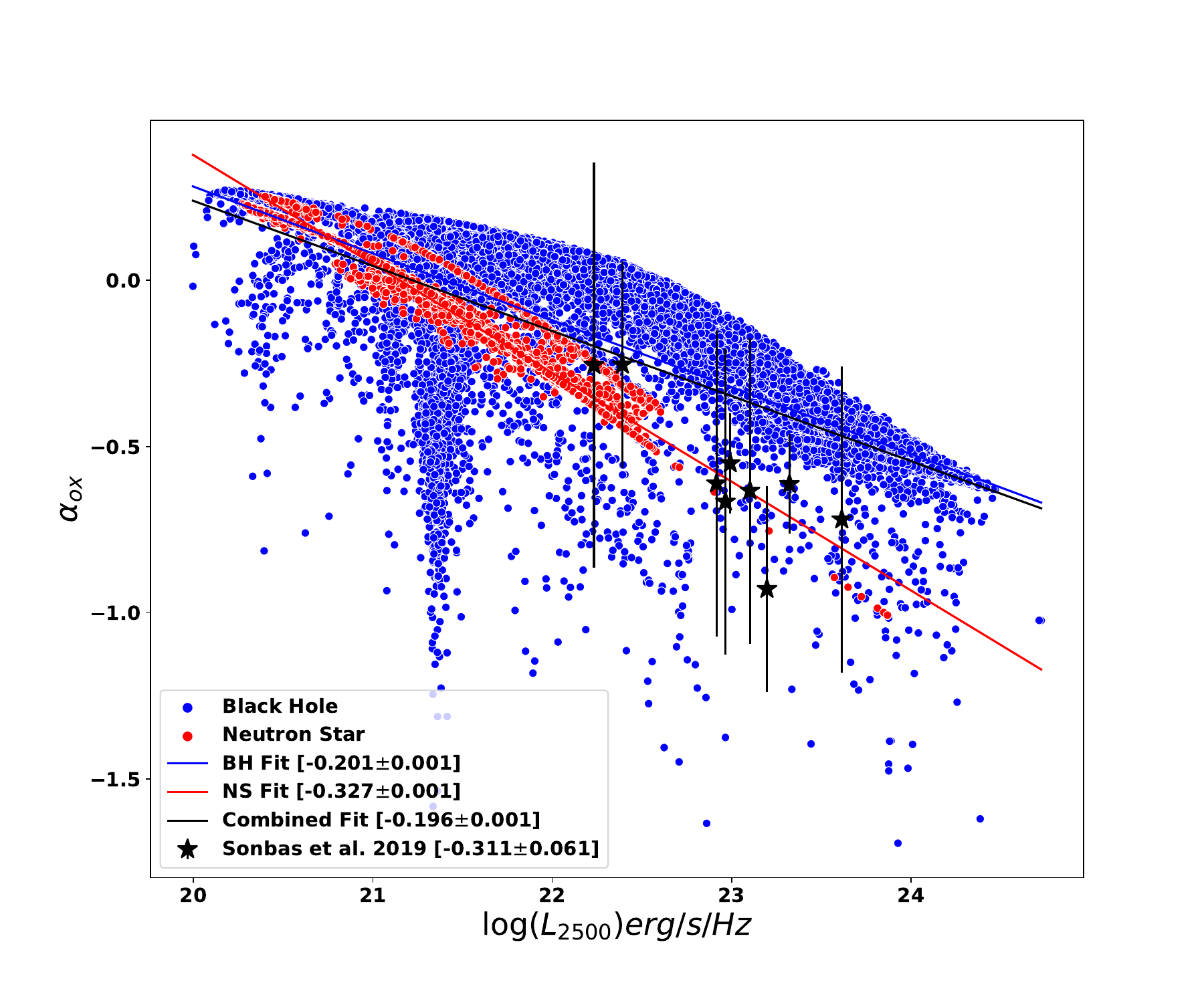}
         \vspace{-0.5cm}
        \label{fig:subfig-b}
    \end{subfigure}
    \hfill
    \begin{subfigure}{0.5\textwidth}
        \centering
        \vspace{-0.6cm}
        \includegraphics[scale=0.3]{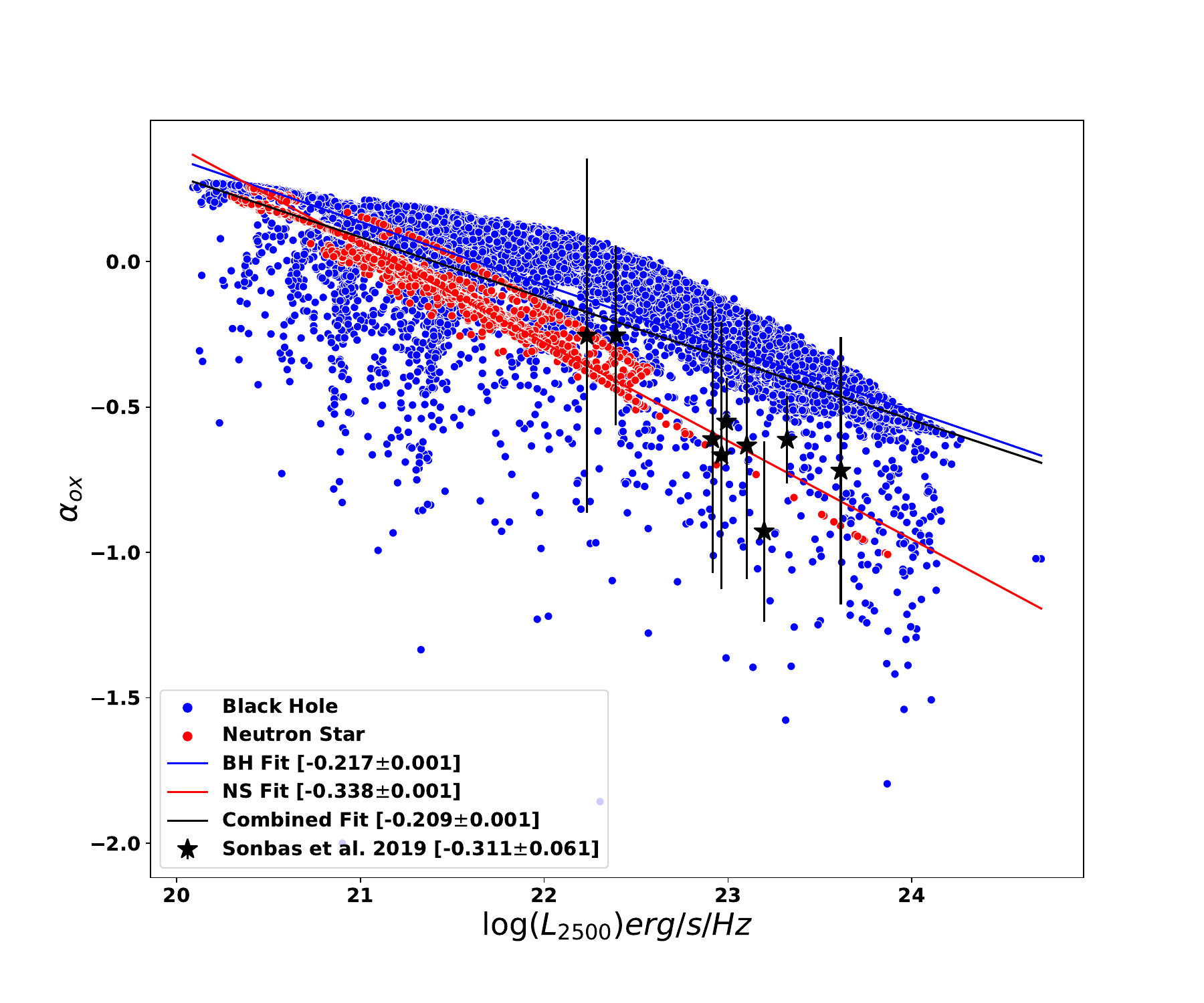}
        \vspace{-0.6cm}
        \label{fig:subfig-c}
    \end{subfigure}
    \caption{$\alpha_{\rm ox}$ index as a function of $L_{\nu,\rm UV}$ (2500 Å) luminosity for our simulated ULX population (BH or NS primaries). The best fit to $\alpha_{\rm ox} - L_{\nu,\rm UV}$ simulated data in our population (black solid line) and observed ULXs data in \citet{sonbas2019evidence} (black points) are also shown. The top, middle and bottom panel correspond to metallicities of $0.025Z_\odot$, $0.20Z_\odot$, and $0.40Z_\odot$, respectively. We also show the $\alpha_{\rm ox} - L_{\nu,\rm UV}$ fits for the BH-ULX and NS-ULX systems separately.}
    \label{figoxbn}
\end{figure}

Fig.~\ref{figoxbn} shows the index $\alpha_{\rm ox}$, defined in Equation (\ref{eq:alphaOX}), as a function of UV luminosity, combining the accretion disk and the companion star, at different metallicities (2.5\%, 20\%, and 40\% of solar metallicity from top-to-bottom) of our simulated ULX population. 
We fit the BH- and NS-ULX simulated populations together and separately with the function, 
\begin{equation}
    \alpha_{\rm OX} = m \log_{10} L_{\nu,\rm UV} +C
\end{equation}
with the slope $m$ and intercept $C$ as free parameters.  The slopes from these fits are shown in Table ~\ref{tab:fit_parameters}. \\

The results presented in Table \ref{tab:fit_parameters} reveal that the slope \(m\) of the \(\alpha_{\text{ox}} - L_{\nu,\rm UV}\) relation varies with metallicity and across different ULX populations. For BH-ULXs, the slope becomes less negative as metallicity increases, indicating that higher metallicities reduce the sensitivity of the X-ray to UV spectral index to UV luminosity. In contrast, NS-ULXs exhibit a relatively stable slope around \(-0.33\) across all metallicities. This consistency
implies that the SED of NS-ULXs is less affected by changes in metallicity compared to BH-ULXs. The combined population shows a similar trend to BH-ULXs, with the slope decreasing slightly with increasing metallicity. Combined slopes are closer to BH-ULX slopes suggests that BH-ULXs may play a slightly more dominant role in determining the overall behavior. This is in line with the simulated BH ULXs outnumbering the NS ULXs in our simulated populations, as evident from Fig.~\ref{fig:Number}.

Also shown in Figure \ref{figoxbn} is the observational data from \citet{sonbas2019evidence}. 
We note that the $\alpha_{\rm ox}$ values for the observed ULXs are  described better by the BH-ULX systems with $f_{\rm out}=0.03$, except possibly for two cases that can be explained by NS-ULX systems. The metallicities of the host galaxies of the ULXs listed in \citet{sonbas2019evidence} are between $0.10Z_\odot$ and $0.36Z_\odot$, with
an average metallicity of $0.22Z_\odot$.  This is approximately the value corresponding to the middle panel of Fig.~\ref{figoxbn}.

Figure \ref{figoxbn} shows that most of the NS-ULX systems in our simulation are in the lower range of UV luminosities compared to the data points from \citet{sonbas2019evidence} with $f_{\rm out}=0.03$. The range of $L_{\nu,\mathrm{UV}}$ is affected by the choice of $f_{\rm out}$. Increasing the $f_{\rm out}$ value leads to high $L_{\nu,\mathrm{UV}}$ (see Section \ref{freeP}). This suggests a potential observational bias, where the observed sample might preferentially include ULXs with higher UV fluxes. This bias could occur because systems with lower UV fluxes are harder to detect and may be underrepresented in the observational sample, leading to a steeper observed slope. Furthermore, our synthetic population includes systems with higher UV luminosity not present in the observational data, indicating that our simulations predict a broader range of UV luminosity for ULX systems.

\subsubsection{Disk vs stellar companion contribution to UV emission}

We calculate UV emission both from the accretion disk and the stellar companion star. We find that a higher percentage of ULX systems are dominated by the disk component in the UV, consistent with observational studies \citep{Tao_2012}. The percentage of disk-dominated ULXs increases with metallicity, starting from 
77\% at the lowest metallicity simulated ($0.005Z_\odot$) to 93\% at the highest metallicity simulated ($1.5Z_\odot$). The percentages of ULXs with disk-dominated UV emission in Fig.~\ref{figoxbn} are 85\%, 82\% and 86\% from the top to the bottom panel, respectively. We also find that the majority of the NS-ULXs are dominated by the disk in the UV, leaving the slope of the companion-dominated UV systems being close to that of the BH-ULXs. We note that the companion stars that do dominate the total UV emission are primarily OB-type stars with very high surface temperature (from 10,000 K to 50,000 K).

\begin{figure}
    \centering
    \includegraphics[scale=0.3]{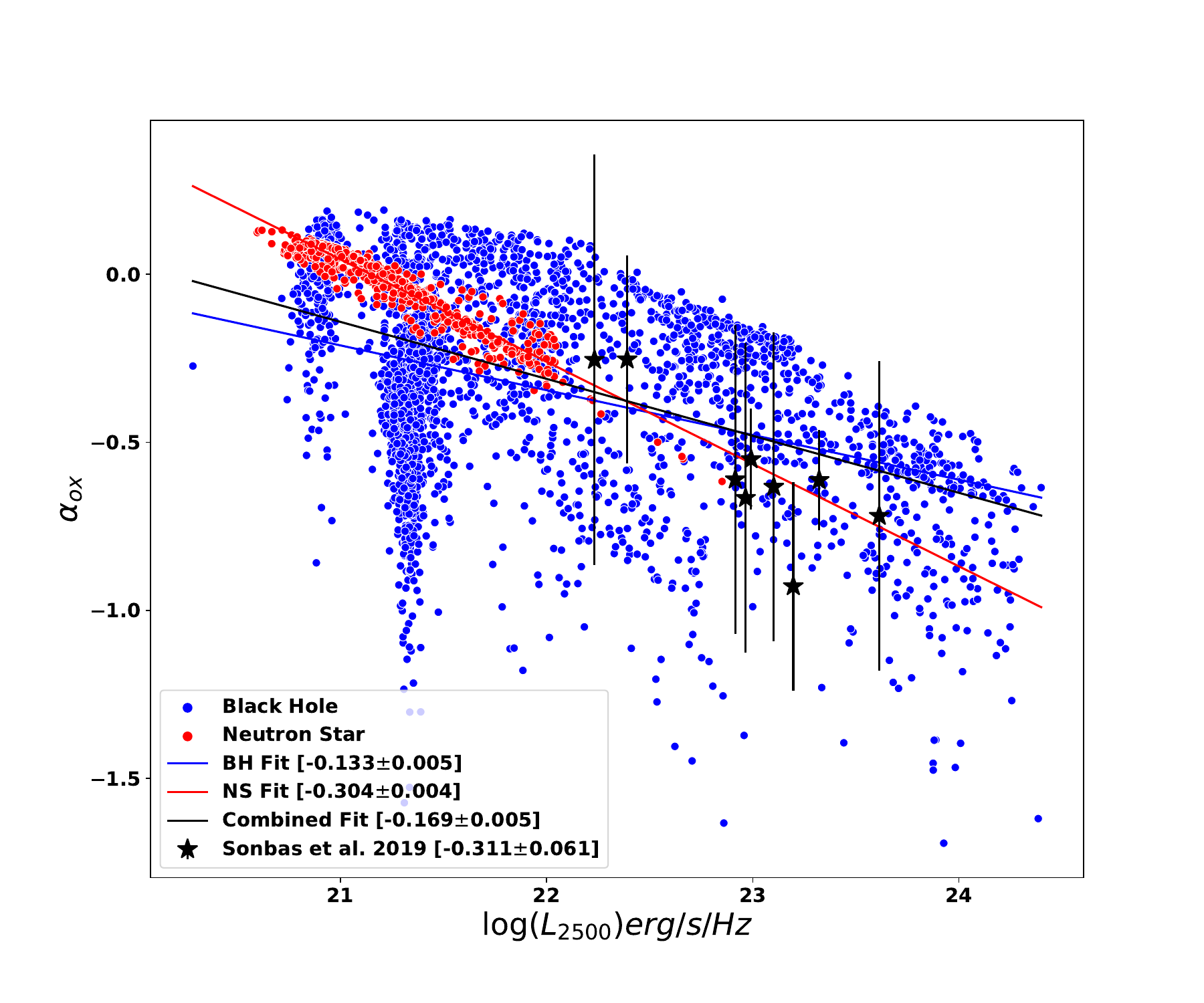}
    \caption{The $\alpha_{\rm ox}$ for donor-star dominated UV emission for ULX systems as a function of UV (2500 Å) luminosity for our simulated ULX population (BH or NS primaries). The best fit to $\alpha_{\rm ox} - L_{\nu,\rm UV}$ simulated data in our population (black solid) at $0.2Z_\odot$ and that for observed ULXs in \citet{sonbas2019evidence} (black points) are also shown.}
    \label{fig:donor}
\end{figure}

Fig.~\ref{fig:donor} shows the $\alpha_{\rm ox}-L_{\nu,\rm UV}$ relation at $0.2Z_\odot$. Note here that this is specifically for the binaries with $L_{\rm UV}$ dominated by the companion. The donor-dominated  BH-ULX systems display a stronger dependence on metallicity, while NS-ULX systems remain relatively unaffected by changes in metallicity (see Table \ref{table:fit_parameters_donor}).  

\begin{table}[h!]
    \centering
    \footnotesize
    \caption{Slope $(m)$ of $\alpha_{\text{ox}}$ vs.\ $L_{\nu,\text{UV}}$.}
    \begin{tabular}{|c|c|c|c|}
        \hline
        Metallicity  & BH-ULX  & NS-ULX  & Combined  \\
        (\% Z$_\odot$)& $m$ & $m$ & $m$ \\
        \hline
        0.5 & $-0.242 \pm 0.000$ & $-0.334 \pm 0.001$ & $-0.221 \pm 0.000$ \\
        2.5 & $-0.234 \pm 0.001$ & $-0.335 \pm 0.001$ & $-0.211 \pm 0.001$ \\
        10 & $-0.225 \pm 0.001$ & $-0.317 \pm 0.001$ & $-0.210 \pm 0.001$ \\
        20 & $-0.201 \pm 0.001$ & $-0.327 \pm 0.001$ & $-0.196 \pm 0.001$ \\
        30 & $-0.210 \pm 0.001$ & $-0.335 \pm 0.001$ & $-0.202 \pm 0.001$ \\
        40 & $-0.217 \pm 0.001$ & $-0.338 \pm 0.001$ & $-0.209 \pm 0.001$ \\
        50 & $-0.221 \pm 0.001$ & $-0.327 \pm 0.001$ & $-0.216 \pm 0.001$ \\
        100 & $-0.213 \pm 0.002$ & $-0.329 \pm 0.001$ & $-0.241 \pm 0.001$ \\
        150&$-0.209\pm 0.00$3&$-0.331\pm 0.001$& $-0.214\pm 0.002$\\
        \hline
    \end{tabular}
    \label{tab:fit_parameters}
    \tablefoot{The slope ($m$) of $\alpha_{\text{ox}}$ vs.\ $L_{\nu,\rm UV}$ at $2500\AA$ and at different metallicities for all ULX systems in our simulations.}
\end{table}

\begin{table}[h!]
\centering
\footnotesize
\caption{Slope $(m)$ of $\alpha_{\text{ox}}$ vs.\ $L_{\nu,\text{UV}}$ donor dominated. }
\begin{tabular}{|c|c|c|c|}
        \hline
Metallicity  & BH-ULX  & NS-ULX  & Combined  \\
        (\% Z$_\odot$) & $m$ & $m$ & $m$ \\
        \hline
        0.5 & $-0.239 \pm 0.002$ & $-0.291 \pm 0.003$ & $-0.224 \pm 0.001$ \\
        2.5 & $-0.229 \pm 0.003$ & $-0.291 \pm 0.004$ & $-0.216 \pm 0.002$ \\
        10 & $-0.163 \pm 0.007$ & $-0.297 \pm 0.007$ & $-0.187 \pm 0.006$ \\
        20 & $-0.133 \pm 0.005$ & $-0.304 \pm 0.004$ & $-0.169 \pm 0.005$ \\
        30 & $-0.193 \pm 0.005$ & $-0.305 \pm 0.003$ & $-0.202 \pm 0.004$ \\
        40 & $-0.232 \pm 0.006$ & $-0.318 \pm 0.003$ & $-0.247 \pm 0.005$ \\
        50 & $-0.272 \pm 0.007$ & $-0.316 \pm 0.004$ & $-0.272 \pm 0.005$ \\
        100 & $-0.343 \pm 0.013$ & $-0.317 \pm 0.005$ & $-0.342 \pm 0.006$ \\
        \hline
\end{tabular}
\label{table:fit_parameters_donor}
\tablefoot{The slope $m$ of $\alpha_{\text{ox}}$ vs.\ $L_{\nu,\rm UV}$ at $2500\AA$ and at different metallicities expressed as a percentage of solar metallicity for systems in which the
UV luminosity is dominated by the donor.}
\end{table}

\subsection{Impact of free parameters \(f_{\rm out}\) and \(\theta_{f}\) on $\alpha_{\rm ox}$}\label{freeP}

 In this model, the key free parameters, \(f_{\rm out}\) and \(\theta_f\), play crucial roles in shaping the reprocessing of disk radiation and the observed spectra. The parameter \(f_{\rm out}\) was determined by the average value in systems explored by \citet{vinokurov2013ultra}. In reality every system will have a different value of $f_{\rm out}$. Unfortunately, we do not have a way to determine the value of \(f_{\rm out}\) for each system; however, we can study the effect of varying this parameter in our population properties.

 The UV emission ($L_{\rm UV}$) is highly sensitive to the value of \(f_{\rm out}\); higher values result in greater UV contribution (see the SED in Fig.~\ref{fig:SEDfout}). Increasing $f_{\rm out}$ makes the values of $\alpha_{\rm ox}$ more negative and the slope of $\alpha_{\rm ox} -L_{\rm \nu,UV}$ relation gets closer to that of \citet{sonbas2019evidence} for the combined population (see Fig.~\ref{fig:alphafout}). With $f_{\rm out}=0.2$, the majority of the observed \citep{sonbas2019evidence} ULXs can be explained by both NS-ULXs and BH-ULXs from our simulations.

 In Fig.~\ref{fig:spectrum}, it can be observed that the frequency corresponding to 2500\,\AA{} does not lie within the flat part of the UV/optical SED for those systems. The wind temperature scales as $f_{\text{out}}^{1/4}$, and decreasing $f_{\text{out}}$ shifts the flat portion of the spectrum toward higher frequencies (or equivalently, shorter wavelengths). Hence, to place 2500\,\AA{} within the flat part of the spectrum, $f_{\text{out}}$ must be decreased further.

 The parameter \(\theta_f\) primarily influences the X-ray emission, making \(\alpha_{\rm ox}\) values more positive with decreasing $\theta_f$, though it has a negligible impact on the slope of the \(\alpha_{\rm ox} - L_{\nu,\rm UV}\) relation for a fixed \(f_{\rm out}\). Decreasing the funnel angle from \(45^{\circ}\) to \(5^{\circ}\) represents a tenfold increase in luminosity. With \(\theta_f = 45^\circ\), the most luminous BH-ULX and NS-ULX at the lowest metallicity have luminosities of \(L_{\rm X} = 1.4 \times 10^{41} \, \text{erg s}^{-1}\) and \(L_{\rm X} = 5.4 \times 10^{39} \, \text{erg s}^{-1}\), respectively, while the most luminous systems have \(L_{\rm X} = 4.2 \times 10^{40} \, \text{erg s}^{-1}\) and \(L_{\rm X} = 5.5 \times 10^{39} \, \text{erg s}^{-1}\) at the highest metallicity simulated.

To fully explore the effects of varying \(\theta_f\), we calculate the funnel angle \(\theta_f\) based on the accretion rate using the model proposed by \cite{king2009masses} in Section \ref{beamsec}.

\subsubsection{Results with beaming and comparative analysis}\label{beamsec}

In the initial phase of our study, we assumed a fixed funnel angle of \(\theta_f = 45^\circ\), allowing us to explore the fundamental properties of ULXs and their X-ray luminosity under this  assumption. However, this assumption does not fully capture the dynamic nature of accretion processes, particularly at high accretion rates where the emission is expected to be highly beamed \citep{king2001ultraluminous}. According to \cite{king2001ultraluminous} and later refined by \cite{king2009masses}, the funnel angle \(\theta_f\) varies with the accretion rate \(\dot{m}\) according to Equation (\ref{eq:beam}). When the accretion rate is high, the luminosity is focused into narrower cones, significantly enhancing the observed luminosity. The beaming factor \(b\) is defined as:
\begin{equation}
b = \frac{\Omega}{4\pi} = 1 - \cos\theta_f \,,
\end{equation}
 where \(\Omega\) is the solid angle of emission. For a conical beam with an opening angle \(\theta_f\), the solid angle is given by
\begin{equation}
\Omega_{\text{total}} = 4\pi (1 - \cos\theta_f) \,.
\end{equation}
 We now calculate the funnel angle \(\theta_f\) based on the accretion rate using the model proposed by \cite{king2009masses}. This equation allows us to compute \(\theta_f\) directly from the beaming factor as
\begin{equation}
\theta_f = \cos^{-1}(1 - b) \,.
\end{equation}
 For \(b = 1\), \(\theta_f = 90^\circ\), and \(r_{\rm ph}\) tends to infinity. To avoid this, we set \(\theta_{f,\rm max} = 87^\circ\), preventing \(r_{\rm ph}\) from becoming infinitely large. With a maximum value of \(\dot{m}\) capped at 1000, the minimum achievable value of \(b\) is \(7 \times 10^{-5}\). However, our simulations show a minimum \(b\) value of \(10^{-4}\). Previous studies have suggested a saturation in the geometric beaming due to high accretion rate, indicating that beyond a certain mass transfer rate (\(\dot{M}\)), further increases in \(\dot{M}\) do not significantly impact the scale height of the accretion disk \citep{lasota2016slimming, wiktorowicz2017origin}. We note that the SCAD model does not put a limit on the geometric thickness of the disk and hence the funnel angle $\theta_f$.
 
 The distribution of \(b\) values from our simulations is illustrated in Fig.~\ref{fig:dist_beam} in Appendix A. We find that the most luminous NS-ULX achieves \(L_{\rm X,b} = 5.3 \times 10^{42}\) erg s$^{-1}$, while the most luminous BH-ULX reaches \(L_{\rm X,b} = 7.5 \times 10^{43}\) erg s$^{-1}$ at the lowest metallicity simulated,  assuming an observer is always looking down the funnel.

\begin{figure}[h]
    \centering
    \includegraphics[width=1.\linewidth]{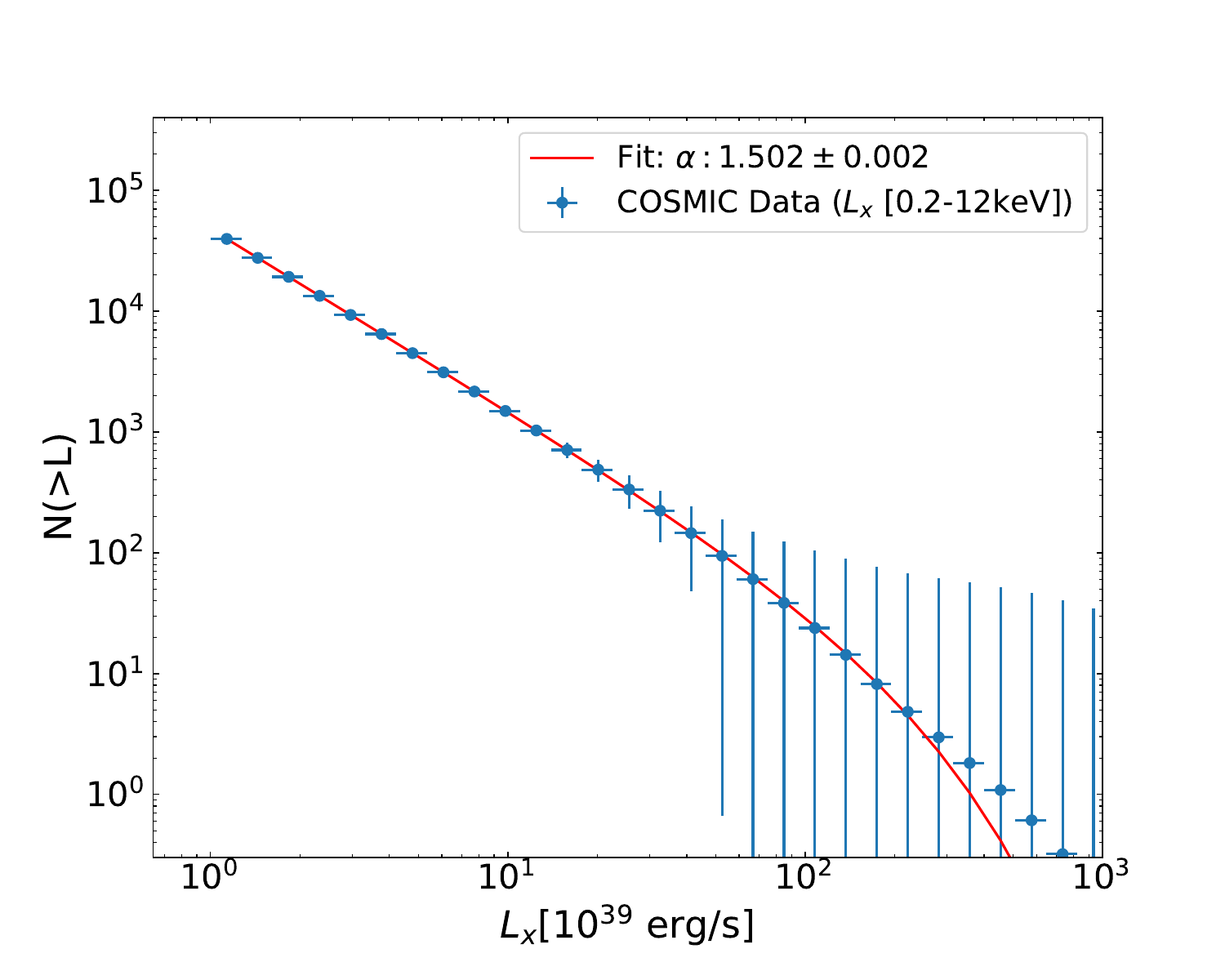}
    \caption{Cumulative XLF of the simulated ULX population at 40\% of $Z_{\odot}$ in the 0.2-12 keV band. }
    \label{fig:XLF}
\end{figure}

 Fig.~\ref{fig:XLF} shows the cumulative XLF of the simulated ULX population at 40\% \(Z_{\odot}\) in the 0.2-12 keV band, incorporating the effects of beaming. The comparison between beamed and non-beamed scenarios reveals that beaming has a significant impact on the luminosity distribution of ULXs. The most luminous sources become even more prominent due to the narrowing of the funnel angle at higher accretion rates, which focuses the emission into a smaller solid angle and therefore increases the observed luminosity if the source is viewed down the disk/wind funnel. The power-law index \(\alpha\) in the beamed scenario at 40\% of \(Z_\odot\) is $1.50$, compared to $1.99$ without beaming. This indicates that beaming increases the number of high-luminosity ULXs, shifting the XLF towards higher luminosities and flattening the slope. This flattening is a direct result of the increased apparent luminosity due to beaming, in systems with high accretion rates.

 \citet{walton20112xmm} constructed XLFs for observed ULXs in spiral and elliptical galaxies, and found \( \alpha = 1.85 \pm 0.11 \) and a steeper index of \( \alpha = 2.5 \pm 0.4 \) for the two galaxy types, respectively. \citet{salvaggio23} studied the XLF of ULXs found in the Cartwheel galaxy and found a slope of \( \alpha = 1.76^{+0.51}_{-0.43} \). Our simulated power-law indices (beamed and non-beamed) fall within the uncertainties of the indices reported for the observed data by these authors.

Table \ref{tab:XLF_beam} presents the \(\alpha\), \(L_s\), and \(\beta\) in Equation (\ref{eq:XLF}) for XLF of ULXs at different metallicities, including the effects of beaming. These parameters highlight the significant impact of beaming on the observed properties of ULX populations. 
The slope \(\alpha\), which characterises the distribution of ULX luminosities, shows a notable difference between the beamed and non-beamed cases. With a constant $\theta_f=45^{\circ}$ (Table \ref{tab:fit_XLF}), the slope \(\alpha\) is consistently steeper across all metallicities, ranging from 1.77 to 2.15, indicating a rapid decline in the number of high-luminosity ULXs. In contrast, with beaming (Table \ref{tab:XLF_beam}), the slope remains nearly constant across all metallicities.

 The dependence of \(N(L_{\rm x})\) on the beaming angle \(\theta_f\) flattens the slope of the XLF and reduces its sensitivity to metallicity. This constant slope suggests that beaming masks the underlying variations in mass distributions across metallicities that would otherwise steepen the XLF, as higher metallicities tend to produce lower-mass compact objects, particularly in case of black holes.

 With beaming, \(L_s\) values are generally higher. For instance, at 0.5\% of \(Z_\odot\), \(L_s = 218 \times 10^{39}\) erg s$^{-1}$, which is much higher than \(L_s = 59.5 \times 10^{39}\) erg s$^{-1}$ with $\theta_f=45^{\circ}$ and at 0.5\% of \(Z_\odot\).

\begin{table}[h!]
\centering
\caption {XLF fit parameters whit beaming included.}
\begin{tabular}{|c|c|c|c|}
    \hline
    Metallicity& slope ($\alpha$) & \(L_s\)        & \(\beta\)       \\
    (\% Z$_\odot$)&&($\times 10^{39}$erg/s)&\\
    \hline
    0.5&1.495\(\pm\)0.003 & 218 \(\pm\) 12       & 1.03 \(\pm\) 0.07       \\
    2.5&1.499\(\pm\) 0.002& 224 \(\pm\) 9      & 1.09 \(\pm\) 0.05       \\
   10 &1.504\(\pm\)0.002 & 224 \(\pm\) 8       & 1.03 \(\pm\) 0.04       \\
    20&1.499\(\pm\) 0.003 & 191 \(\pm\) 6       & 1.00 \(\pm\) 0.04       \\
   30 &1.499\(\pm\)0.002 & 170 \(\pm\) 6       & 1.05 \(\pm\) 0.05        \\
   40 & 1.502\(\pm\) 0.002& 191 \(\pm\) 6       & 1.04 \(\pm\) 0.04       \\
   50&1.500\(\pm\) 0.002& 161 \(\pm\) 5       & 1.17 \(\pm\) 0.05        \\
   100 &1.514 \(\pm\) 0.001 & 110 \(\pm\) 2       & 1.45 \(\pm\) 0.04        \\
   150 &1.504\(\pm\) 0.002& 107 \(\pm\) 2       & 1.49 \(\pm\) 0.02       \\
        \hline

\end{tabular}
\label{tab:XLF_beam}
\tablefoot{Values of XLF parameters $\alpha$, \(L_s\), and \(\beta\) with their respective uncertainties when beaming is included.}
\end{table}

\begin{figure}
    \centering
    \begin{subfigure}{0.5\textwidth}
        \centering
        \includegraphics[scale=0.3]{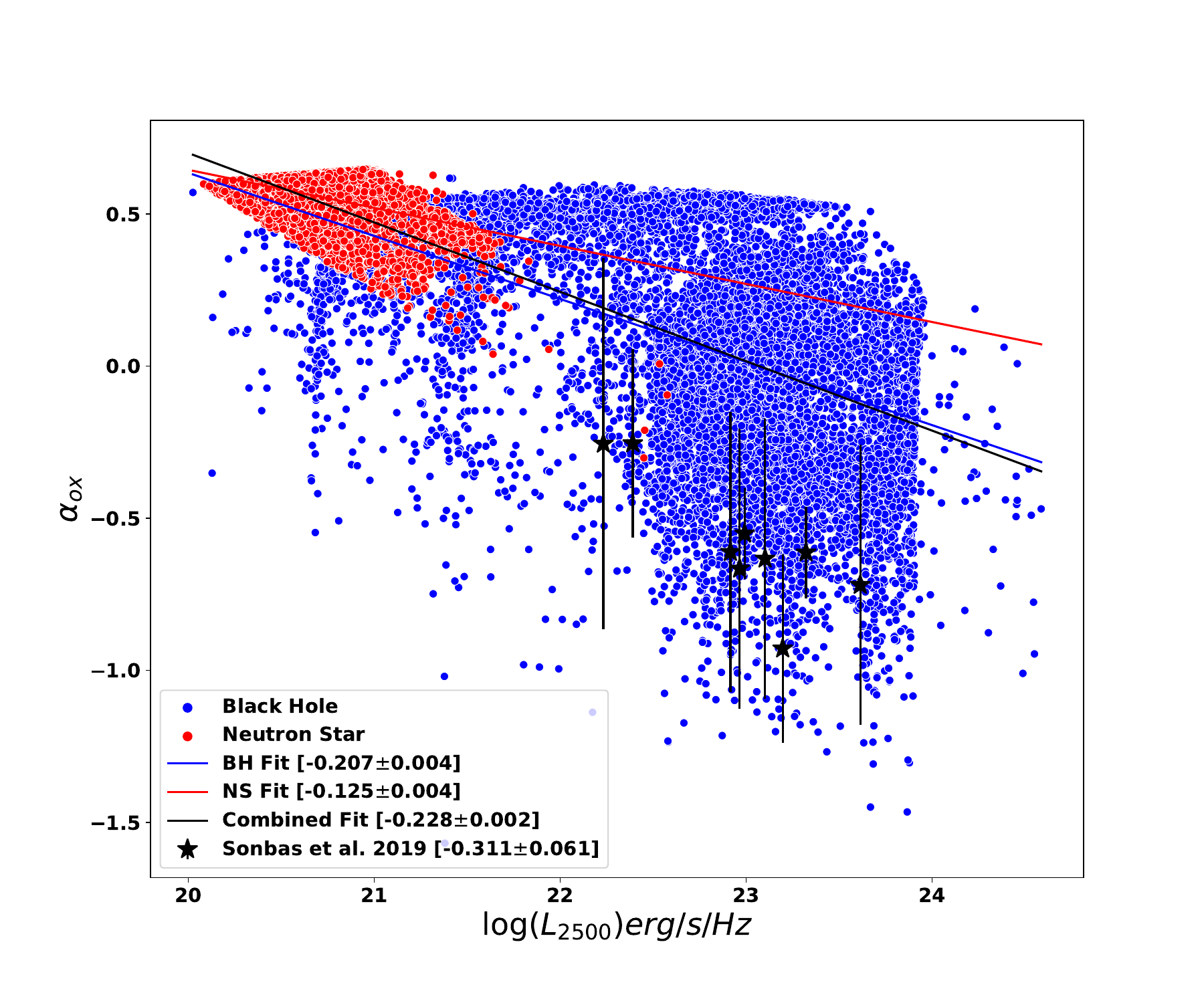}
        \vspace{-0.6cm}
        \label{fig:subfig-a}
    \end{subfigure}
    \hfill
    \begin{subfigure}{0.5\textwidth}
        \centering
        \vspace{-0.6cm}
        \includegraphics[scale=0.3]{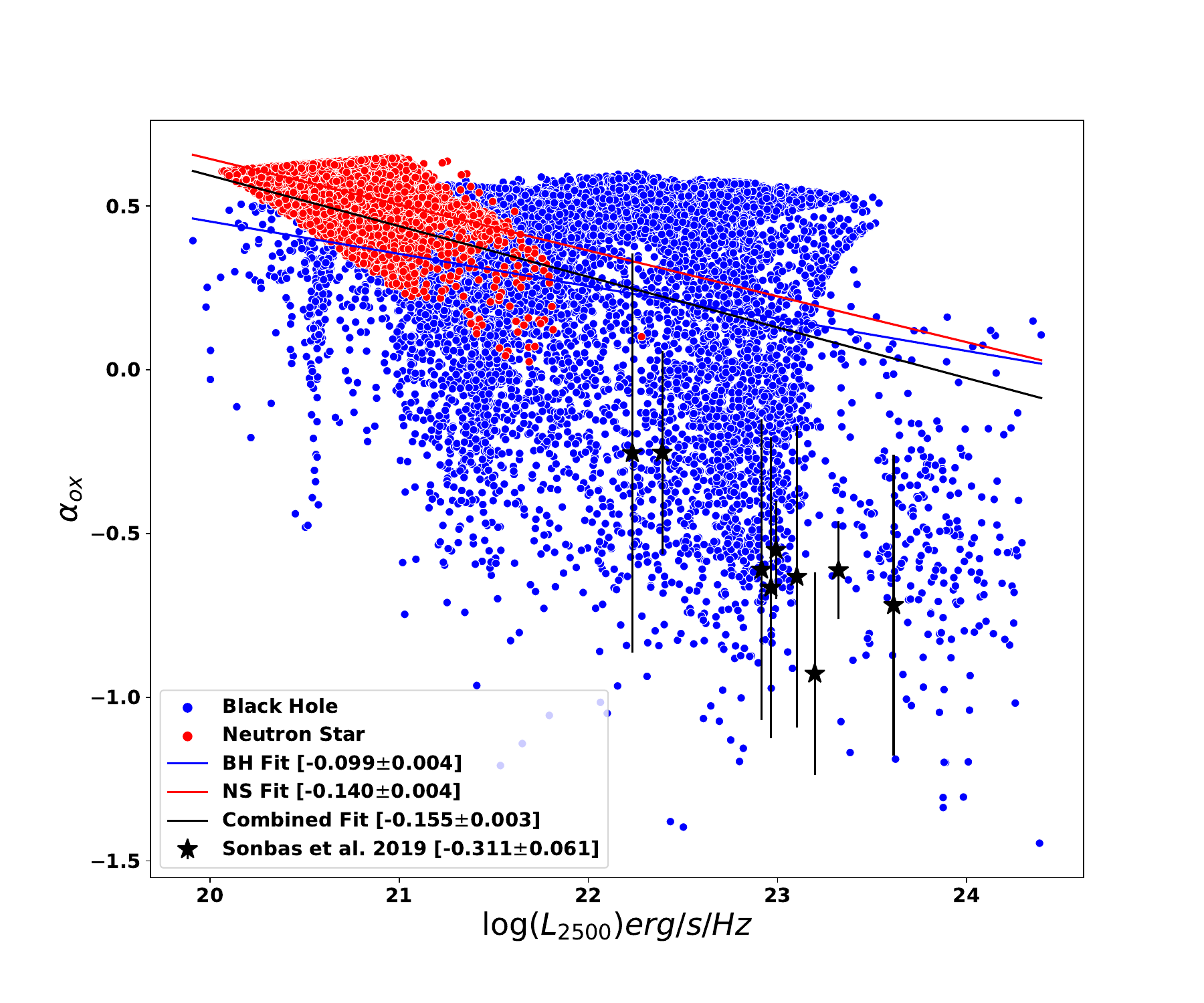}
         \vspace{-0.5cm}
        \label{fig:subfig-b}
    \end{subfigure}
    \hfill
    \begin{subfigure}{0.5\textwidth}
        \centering
        \vspace{-0.6cm}
        \includegraphics[scale=0.3]{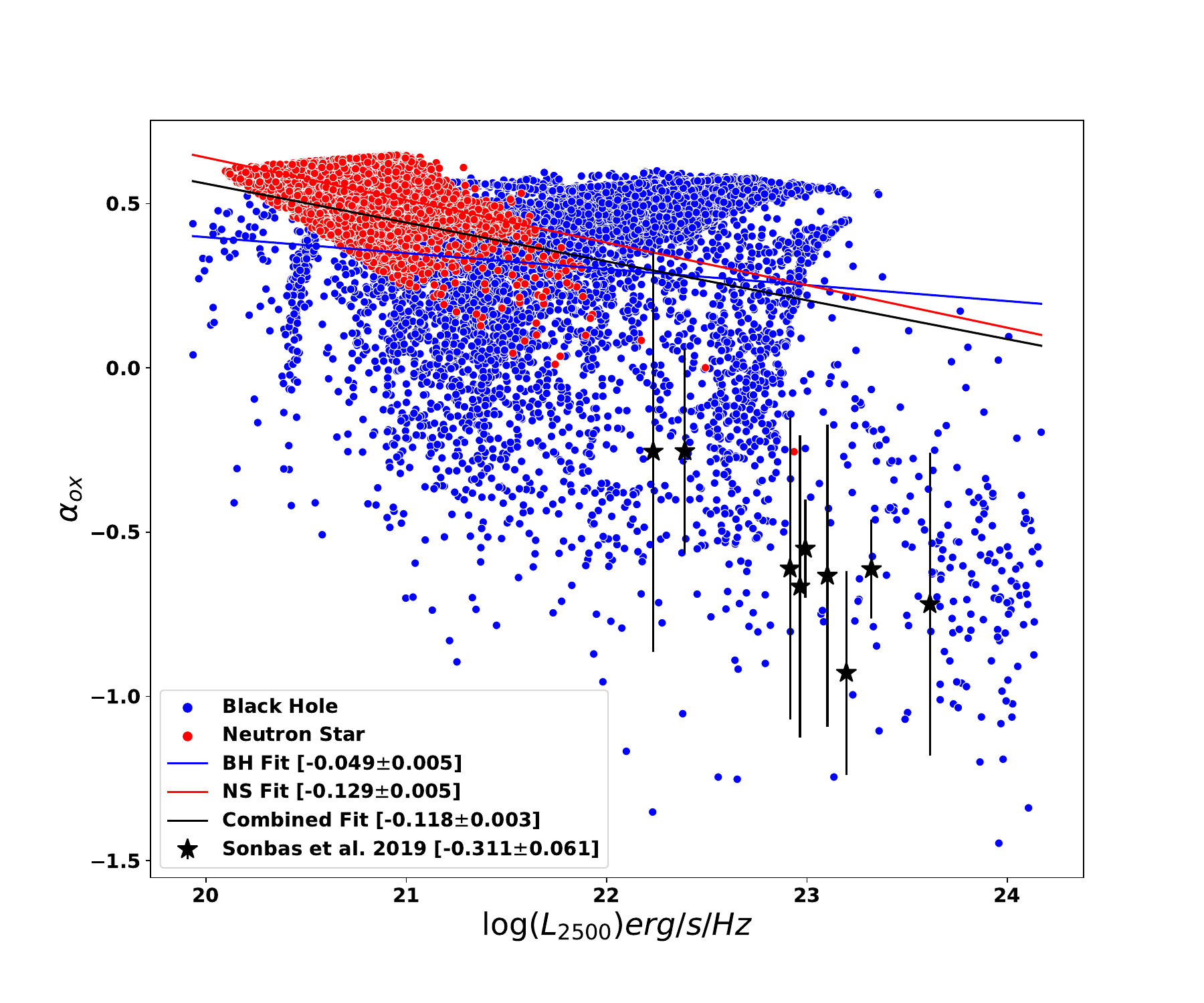}
        \vspace{-0.6cm}
        \label{fig:subfig-c}
    \end{subfigure}
    \caption{$\alpha_{\rm ox}$ index as a function of $L_{\nu,\rm UV}$ (2500 Å) luminosity for our simulated ULX population (BH or NS primaries) with beaming included in the calculation. The best fit to $\alpha_{\rm ox} - L_{\nu,\rm UV}$ simulated data in our population (black solid line) and observed ULXs data in \citet{sonbas2019evidence} (black points) are also shown. The top, middle and bottom panel correspond to metallicities of $0.25Z_\odot$, $0.2Z_\odot$, and $0.4Z_\odot$, respectively. }
   \label{figoxbn_beam}
\end{figure}

The introduction of beaming also impacts the optical-X-ray spectral index, \(\alpha_{\rm ox}\). Fig. \ref{figoxbn_beam} presents \(\alpha_{\rm ox}\) as a function of \(L_{\nu,\rm UV}\) (2500 Å luminosity) for our simulated ULX population, including beaming. The slopes (\(m\)) of the \(\alpha_{\text{ox}}\) vs. \(L_{\nu,\rm UV}\) relation differ significantly between the beamed (Table \ref{tab:fit_beam}) and non-beamed (Table \ref{tab:fit_parameters}) at high metallicity, highlighting the substantial impact of beaming on the relative X-ray and UV emissions of ULXs.

Including beaming results in high-beamed systems exhibiting more positive $\alpha_{\rm ox}$ values compared to the observational data, leading to more data points lying above those of \citet{sonbas2019evidence}. In general, introducing beaming reduces the magnitude of the slope \(m\), particularly at higher metallicities. This suggests that at higher metallicities beaming results in a larger number of systems exhibiting more focused emission, leading to a flattening of the \(\alpha_{\text{ox}} - L_{\nu,\rm UV}\) relation. As a result, metallicity plays a diminished role in determining the observed correlation, with beaming effects becoming the dominant factor influencing the properties of these ULX systems. Furthermore, we notice that inclusion of beaming leads to a large scatter in the \(\alpha_{\text{ox}} - L_{\nu,\rm UV}\) plane, and a linear fit may not be appropriate as can be inferred from near-flat slope of the fits.

\begin{table}[h!]
    \centering
    \footnotesize
    \caption{ Slope of $\alpha_{\text{ox}}$ vs.\ $L_{\nu,\rm UV}$ with beaming included.}
    \begin{tabular}{|c|c|}
        \hline
        Metallicity  &  Combined  \\
        (\% Z$_\odot$)& $m$ \\
        \hline
        0.5 & $-0.217 \pm 0.002$\\
        2.5 & $-0.228 \pm 0.002$ \\
        10 & $-0.189 \pm 0.003$ \\
        20 & $-0.155 \pm 0.003$ \\
        30 & $-0.154 \pm 0.003$ \\
        40 & $-0.118 \pm 0.003$ \\
        50 & $-0.123 \pm 0.003$ \\
        100 & $-0.079 \pm 0.004$ \\
        150 & $-0.122 \pm 0.003$ \\
        \hline
    \end{tabular}
    \label{tab:fit_beam}
    \tablefoot{The slope $m$ of $\alpha_{\text{ox}}$ vs.\ $L_{\nu,\rm UV}$ at $2500\AA$ and at different metallicities for all ULX systems in our simulations with beaming included.}
\end{table}

\subsubsection{\bf Neutron star ULXs}

\begin{figure}
    \centering
    \begin{subfigure}{0.5\textwidth}
        \centering
        \includegraphics[width=1.\linewidth]{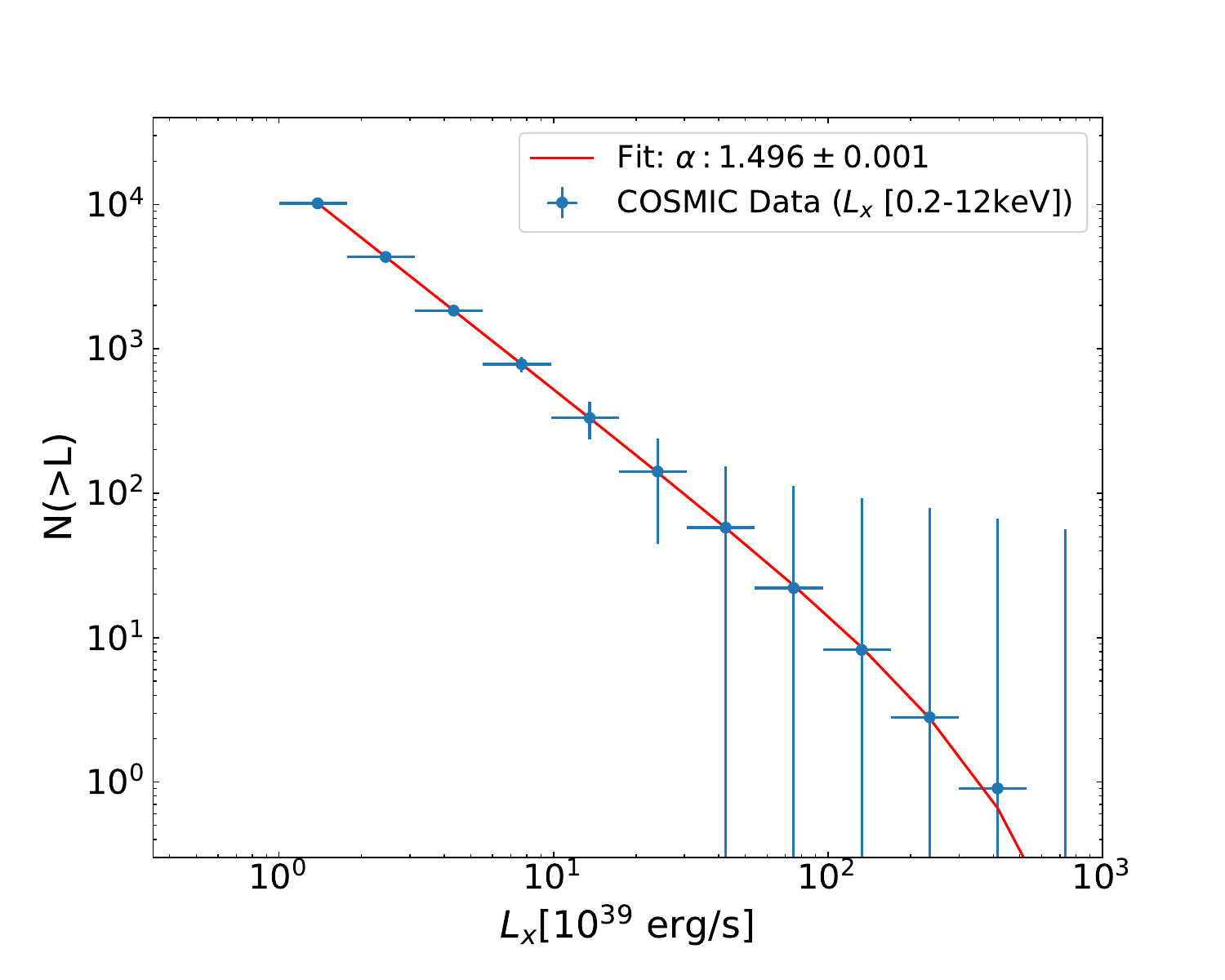}
        \vspace{-0.6cm}
        \label{fig:subfig-a}
    \end{subfigure}
    \hfill
    \caption{ Cumulative XLF of the simulated NS-ULX population at 40\% of $Z_{\odot}$ in the 0.2-12 keV band including beaming. }
    \label{fig:NS_XLF}
\end{figure}

Pulsations found in some ULX systems have provided compelling evidence that at least a fraction of ULXs are powered by NS \citep{bachetti2014ultraluminous,furst2016discovery}. In Fig.~\ref{fig:NS_XLF}, we present the XLF of our simulated NS-ULX population at $Z=40\% Z_{\odot}$. Our simulations reveal that the XLF of NS-ULXs have a slope similar to the total ULX population, though with slightly higher $L_{\rm s}$ and $\beta$, with values in Fig.~\ref{fig:NS_XLF} given by $381\pm29\times 10^{39}$erg s$^{-1}$  and $1.3\pm 0.1$ respectively.

We find that the average value of $\theta_f \approx 22^{\circ}$ for all NS ULXs at $Z=40\% Z_{\odot}$, which corresponds to {$1/b\approx 13$}. Therefore, a beaming factor of at least 10 is required to achieve extreme luminosities ($L_x > 10^{40}$ erg/s) observed in some NS-ULXs. However, high pulsed fractions observed in some pulsating ULXs  \citep{2022ApJ...937..125B} indicate against strong beaming in these systems \citep{2021MNRAS.501.2424M}. 
Furthermore, \citet{2022ApJ...937..125B} showed that the extreme luminosity of M82 X-2 can be explained by super-Eddington accretion and high mass transfer rates on to a highly magnetised NS.

 The omission of magnetic field in our model presents a limitation. Including magnetic field can significantly affect the accretion geometry and emission properties on the NS-ULX systems \citep[e.g.,][]{middleton2019magnetic,2021MNRAS.501.2424M,2022ApJ...937..125B}. Incorporating these effects is, however, beyond the scope of this work.

\subsubsection{\bf Observable ULX population}

 Beamed emission from a ULX is observable if the observer is within $\theta_f$. To study what fraction of our ULX population is observable, we assign random viewing angle $(\beta)$ to all ULXs using the distribution $P(\beta)=\sin(\beta)$. The cumulative distribution function (CDF) of \( P(\beta) \) is then
\begin{equation} 
C(\beta) = \int_0^\beta P(\beta') \, d\beta' = \int_0^\beta \frac{1}{2} \sin\beta' \, d\beta' = \frac{1}{2}(1 - \cos\beta).
\end{equation}
Here \( C(\beta) \) ranges from \( 0 \) (at \( \beta = 0 \)) to \( 1 \) (at \( \beta = \pi \)) and $\beta = 0$ a face-on case, and $\beta = \pi/2$ is an edge-on case. Therefore,
\begin{equation}
\beta = \arccos(1 - 2C).
\label{beta_v}
\end{equation}
{We use Eq.~(\ref{beta_v}) to calculate $\beta$, where $C$ is randomly sampled from a uniform distribution in $[0, 1]$. A ULX is observable if
}
\[
\theta_f < \beta \quad \text{or} \quad \theta_f > \pi - \beta.
\]

\begin{table}[h!]
\caption{Observable ULX population.}
\centering
\begin{tabular}{|c|c|c|c|}
\hline
Metallicity (\%) & NS-ULX  & BH-ULX  & All ULX \\
$(\%~Z_{\odot})$&(\%)& (\%)&(\%)\\
\hline
0.5  & 10.04 & 42.62 & 31.50 \\ 
2.5  & 6.27  & 43.00 & 29.06 \\ 
10   & 7.61  & 58.61 & 40.88 \\ 
20   & 8.29  & 64.82 & 43.24 \\ 
30   & 8.59  & 67.22 & 43.52 \\ 
40   & 7.58  & 66.92 & 41.09 \\
50   & 8.26  & 67.08 & 38.12 \\ 
100  & 8.56  & 74.32 & 40.95 \\ 
150  & 8.12  & 78.20 & 39.81 \\ \hline
\end{tabular}
\label{tab:obse}
\tablefoot{The percentages of observable NS-ULX, BH-ULX, and combined ULX population at different metallicities.}
\end{table}
 
Table \ref{tab:obse} shows the percentage of ULX population observable after viewing angle correction. A relatively small fraction ($\lesssim 10\%$) of the NS ULXs are observable as compared to the BH ULXs. Notably, the NS-ULX fraction remains largely independent of metallicity while the BH ULXs exhibit a much higher fraction, ranging from 43\% at 0.5\% \(Z_\odot\) to 78\% at 150\% \(Z_\odot\). This difference in the observability of the NS ULXs and BH ULXs can be understood from their beaming distributions shown in Fig.~\ref{fig:obse_dist_beam}. 

\begin{figure}[H]
    \centering
    \includegraphics[width=1.\linewidth]{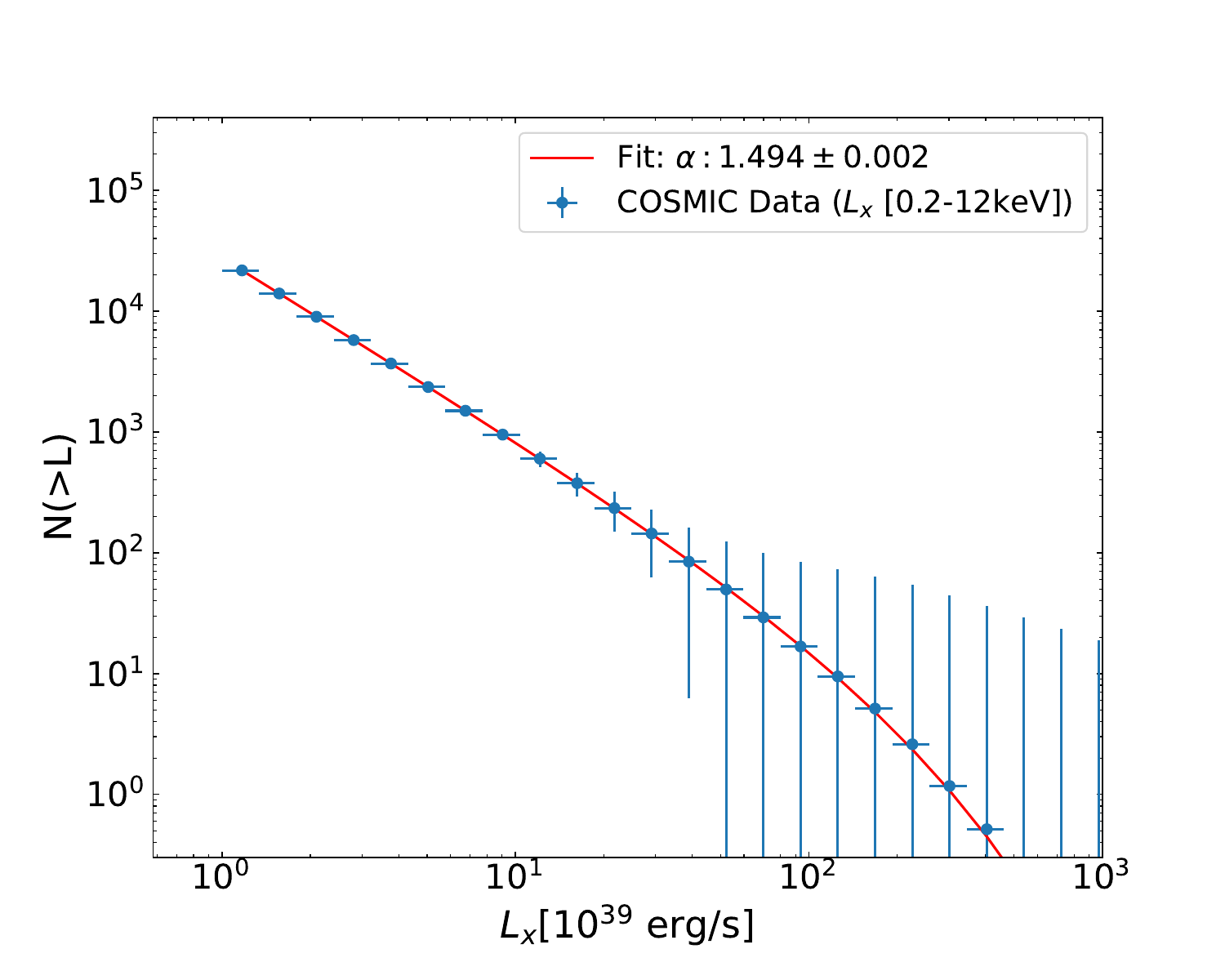}
    \caption{ The cumulative XLF of the simulated observable ULXs at 40\% of $Z_{\odot}$ in the 0.2-12 keV band including beaming.}
    \label{fig:LF_obse}
\end{figure}

 Although the fraction of observable ULXs remains below 44\% across all metallicities, their properties are not substantially different from those of the entire simulated population. Fig. \ref{fig:LF_obse} displays the luminosity function of the observable ULXs at \(40\%~Z_{\odot}\), while Fig.~\ref{fig:obse_ox} presents their \(\alpha_{\rm ox}-L_{\nu,\rm UV}\) relation. The observable population shows a slightly steeper \(\alpha_{\rm ox}-L_{\nu,\rm UV}\) slope. It is important to note, however, that our analysis of the observable population does not account for any plausible selection effects of the detector.

\begin{figure}
    \centering
    \includegraphics[scale=0.33]{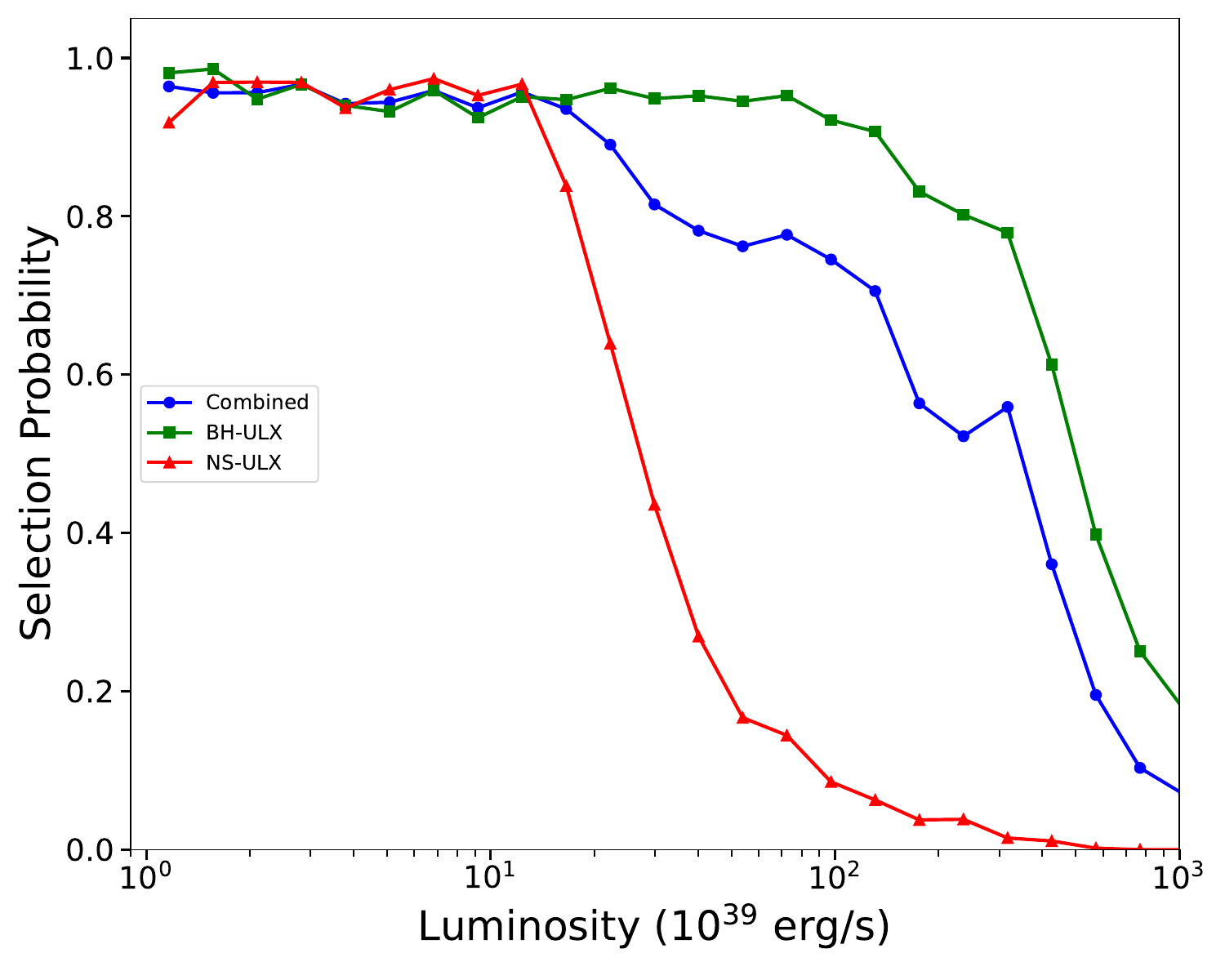}
    \caption{The selection probability, a ratio of the observable to the total number of all ULX systems (blue), BH-ULX (green), and NS-ULX (red), as a function of X-ray luminosity at $40\%~Z_{\odot}$.}
    \label{fig:selection_probability}
\end{figure}

\begin{figure}
    \centering
    \begin{subfigure}{0.5\textwidth}
        \centering
        \includegraphics[scale=0.3]{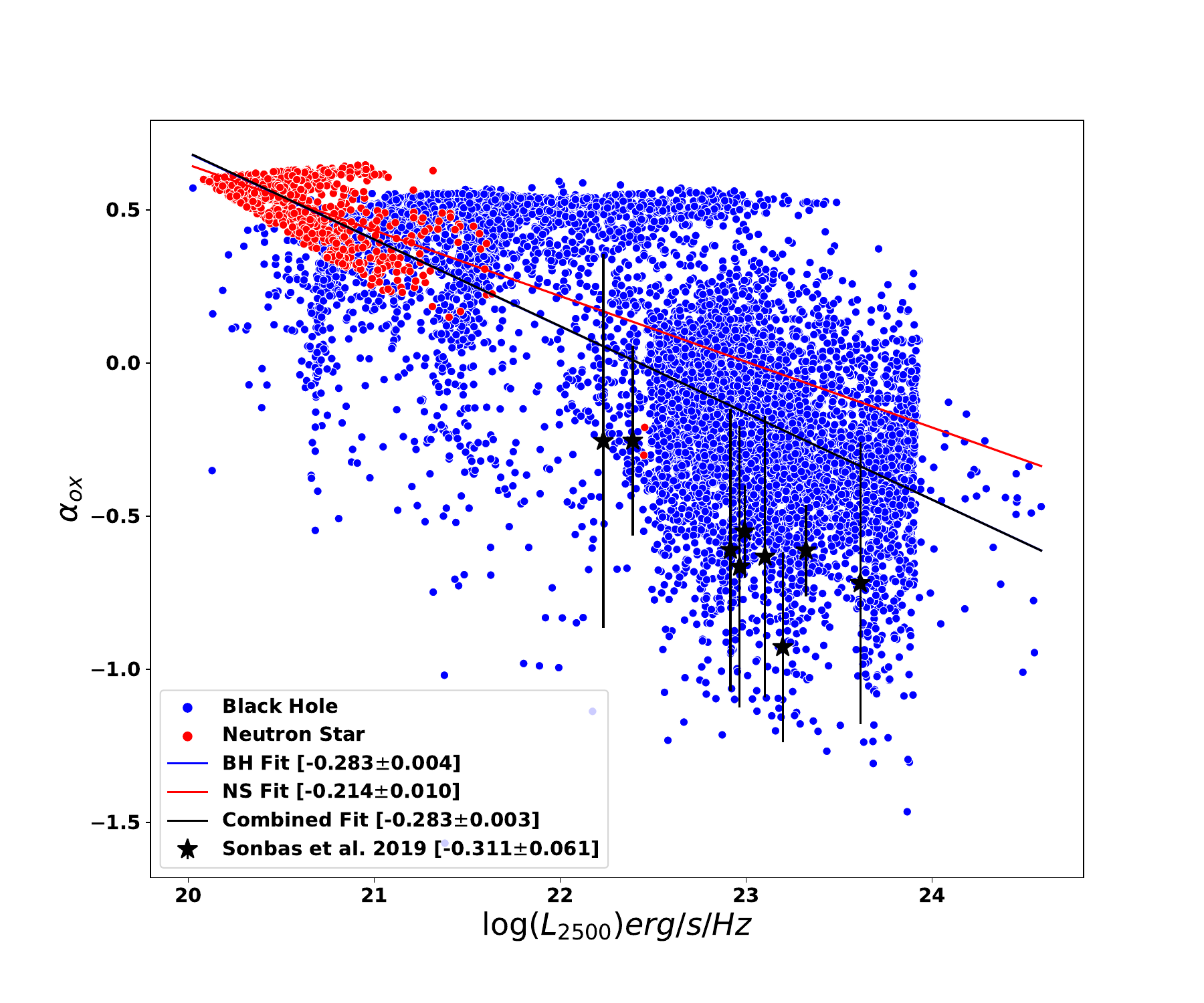}
        \vspace{-0.6cm}
        \label{fig:subfig-a}
    \end{subfigure}
    \hfill
    \begin{subfigure}{0.5\textwidth}
        \centering
        \vspace{-0.6cm}
        \includegraphics[scale=0.3]{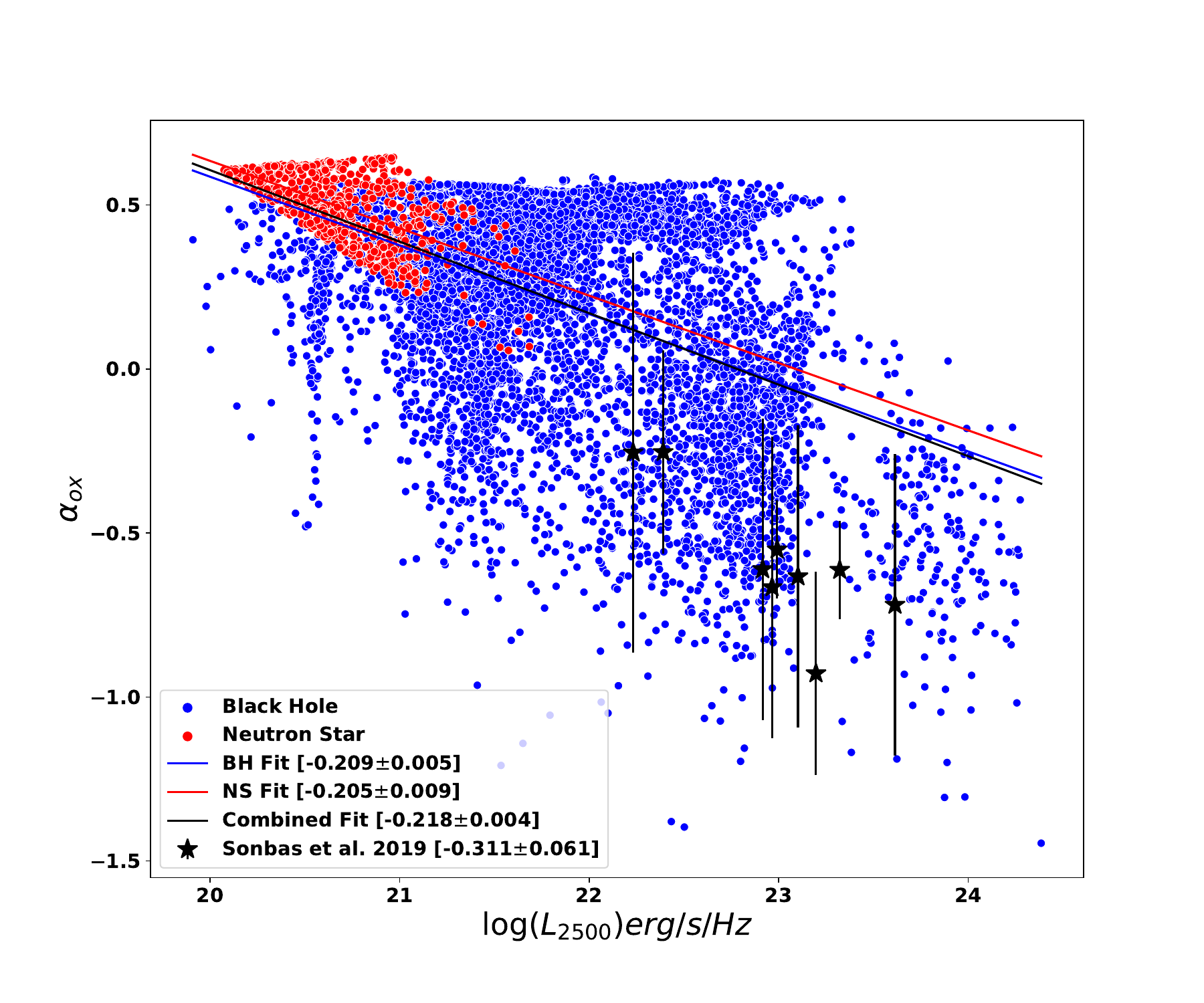}
         \vspace{-0.5cm}
        \label{fig:subfig-b}
    \end{subfigure}
    \hfill
    \begin{subfigure}{0.5\textwidth}
        \centering
        \vspace{-0.6cm}
        \includegraphics[scale=0.3]{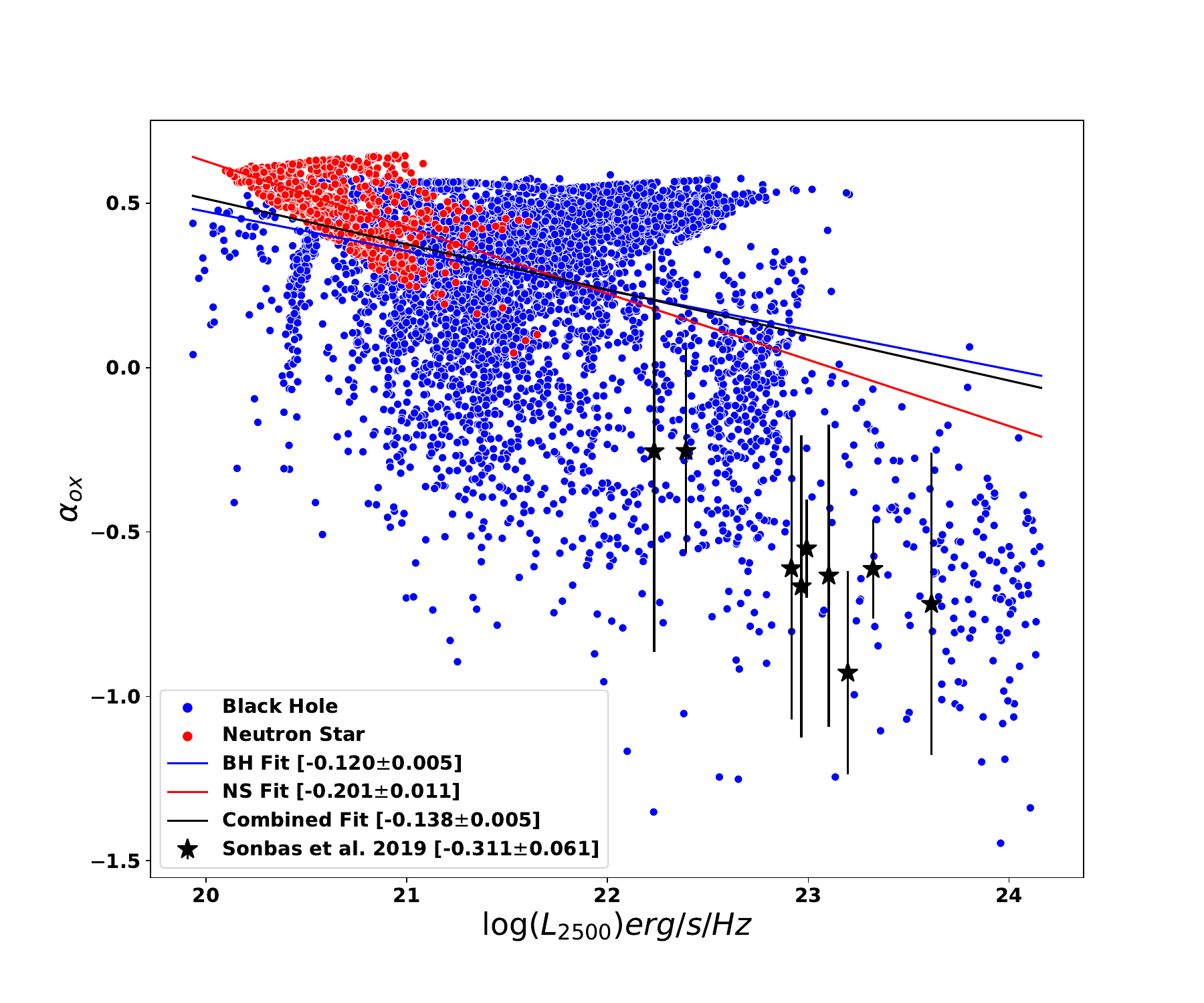}
        \vspace{-0.6cm}
        \label{fig:subfig-c}
    \end{subfigure}
    \caption{ $\alpha_{\rm ox}$ vs.\ $L_{\nu,\rm UV}$ (2500 Å) for our simulated observable ULX population (BH or NS primaries) with beaming included in the calculation. The best fit to the simulated data (black solid line) and observed ULXs in \citet{sonbas2019evidence} (black points) are also shown. The top, middle and bottom panel correspond to metallicities of $0.25Z_\odot$, $0.2Z_\odot$, and $0.4Z_\odot$, respectively. }
    \label{fig:obse_ox}
\end{figure}

  To investigate the effect of selection criteria on the XLF, we computed the selection probability as a function of luminosity. To calculate this probability we divide the luminosity range into logarithmically spaced bins and computed the total number of systems and the number of observable systems in each bin. The selection probability is then defined as the ratio of the observable to the total number of systems. Figure \ref{fig:selection_probability} shows the selection probability for population in Figure \ref{fig:XLF} and the observable population in Fig.~\ref{fig:LF_obse}. 
 The results reveal that the probability remains approximately flat between $10^{39}$ and $2\times10^{40}~\mathrm{erg~s^{-1}}$ and this is where most of the ULX systems are located. 

 The invariance of the XLF slope is primarily driven by the dominance of ULX systems within the flat-probability range in Fig.~\ref{fig:selection_probability}. In this region, the probability of making an observation is high due to low beaming resulting in wider opening angle. At high luminosities ($L > 2 \times 10^{40}$~erg~s$^{-1}$), the NS-ULX emission becomes increasingly beamed into narrow cones, significantly reducing the likelihood of observation and resulting in a drop in the total probability. In contrast, the BH-ULX selection probability remains constant below luminosities of $L \sim 10^{41}$ erg s$^{-1}$, where beaming is unimportant.

 The selection process does not introduce significant luminosity-dependent biases in this range, allowing the intrinsic XLF properties to be preserved in the observed sample. It is also important to remind the reader that the viewing angle is randomly assigned from a $\sin(\beta)$ distribution. Hence, the luminosity does not depend on the angle $\beta$.

\section{Summary and conclusions}
\label{summary_section}
We summarise our results below.

\begin{enumerate}
\item In this study, we used the \textsc{cosmic} population synthesis code to simulate the evolution of binary systems and investigate the relationship between UV and X-ray emission during the ULX phase.

\item We have adopted the SCAD model by \citet{vinokurov2013ultra} to calculate SEDs of our simulated ULX population from the accretion disk. While X-rays are emitted solely by the accretion disk, the UV emission is a combination of heated wind from the disk and the surface emission from the companion star.
 
\item Our findings indicate that the number of ULXs decreases with increasing metallicity, consistent with observational data. We found that BH-ULXs decrease at a faster rate than NS-ULXs with increasing metallicity, and obtained a power law index of \( \alpha = 0.16 \pm 0.01 \) for the combined population of BH- and NS-ULXs, which is consistent with values reported in literature. The decreasing number of ULXs with increasing metallicity has significant implications for our understanding of ULX populations in different galactic environments. In metal-poor galaxies, we expect a higher number of ULXs, which can serve as important probes of early galaxy evolution.
    
\item 
Luminosity Function: The XLF slopes of our simulated ULX population vary with metallicity, with lower metallicity populations exhibiting a relative excess of high-luminosity ULXs. Without beaming, the slope ranges from \( \alpha = 1.77 \pm 0.01 \) at 0.5\% solar metallicity to \( \alpha = 2.15 \pm 0.38 \) at 1.5 times solar metallicity. The inclusion of beaming flattens the XLF slope across all metallicities, reflecting the boosted luminosity due to the beaming effect. Our simulated slopes are consistent with observational data from \citet{walton20112xmm} and \citet{salvaggio23}, who reported similar values for ULXs in different types of galaxies.

\item The optical--X-ray spectral index \(\alpha_{\rm ox}\) showed a significant anti-correlation with UV luminosity across different metallicities. This relation is found using both the disk and the stellar companion contributions to the UV emission. Our results suggest that UV emission in ULXs is predominantly disk-dominated, with the percentage of disk-dominated ULXs increasing as metallicity rises. The slope of the \(\alpha_\mathrm{ox} - L_{\nu,\rm UV}\) relation for BH-ULXs changes with metallicity, becoming steeper at higher metallicities, while the slope for NS-ULXs remains relatively constant. With beaming included, the slopes of the \(\alpha_{\rm ox} - L_{\nu,\rm UV}\) relation become significantly flatter and remain relatively constant with metallicity.

\item 
{For a constant $\theta_f$, our results suggest that the majority of the ULXs with unique optical counterparts selected by \citet{sonbas2019evidence} are likely powered by black holes when $f_{\rm out}=0.03$, however, for high $f_{\rm out}>0.1$ more NS-ULX systems are found to be compatible with the observational data. Including beaming in our calculation resulted in high $\alpha_{\rm ox}$ values compared to \citet{sonbas2019evidence} data points.  }

\item Our results support the hypothesis that stellar-mass BHs and NSs, undergoing supercritical accretion, can account for the observed properties of ULXs without necessitating the presence of IMBHs.

\item  While the X-ray emission remains unaffected, higher wind velocity reduces UV emission, causing the $\alpha_{\rm ox}$ values to become more positive.

\item  Due to their high beaming, NS-ULX systems produce narrow emission cones, which significantly reduce their likelihood of being observed.

\end{enumerate}

 We note that \textsc{cosmic} has a few limitations in simulating the evolution of binary populations. For example, the relationship between remnant mass and initial mass or carbon-oxygen core mass is not always straightforward. Other theoretical models have shown that some massive stars directly implode into black holes without an explosion \citep{10.1093/mnras/staa3029}, which \textsc{cosmic} does not fully account for. Unlike advanced stellar evolution code like Modules for Experiments in Stellar Astrophysics \citep[MESA;][]{Paxton_2010} that provides detailed simulations of the physical processes governing the life cycle of stars, \textsc{cosmic} is a rapid population synthesis code that relies on simplified models for stellar and binary evolution, introducing systematic biases in the resulting binary populations \citep{2021A&A...650A.107M, Gallegos-Garcia_2021}. These limitations can be addressed by population synthesis codes such as \textsc{POSYDON} \citep{Fragos_2023}, which incorporate a detailed MESA code. However, current versions of these codes are constrained to a limited range of metallicity and are not yet suitable for studies like ours, where metallicity is varied over a wide range.

 We also note a few limitations of our radiation calculations. The SCAD model has been developed for BH-ULX systems. Its application to NS-ULX systems in our study has limitations. One significant limitation is that we did not account for the effects of magnetic fields in NS-ULX systems. Magnetic fields can play a critical role in the dynamics of the accretion disk, potentially leading to disk truncation and the occurrence of pulsations, both of which are characteristic features in NS-ULXs \citep{middleton2019magnetic}. Calculations by \citet{2021MNRAS.501.2424M} suggest that strong magnetic fields ($B > 10^{13}$~G) in systems like M82 X-2 facilitate super-Eddington accretion and reduce reliance on beaming to achieve high luminosities. Similarly, \citet{2022ApJ...937..125B} argue that moderate beaming combined with accretion funnel effects and magnetic field interactions better explains the observed properties of pulsating ULXs. Additionally, the irradiation of the donor star by the accretion disk, which can increase UV emission \citep{motch2014mass}, was not considered, potentially affecting the accuracy of the \(\alpha_{\rm ox}\) vs. \(L_{\nu,\rm UV}\) relation. Our future studies will aim to improve on some of those limitations.

 
\begin{acknowledgements}
We are grateful to the referee for helpful comments that have improved this paper. We would like to thank Kalvir Dhuga, Eda Sonbas and Andrzej Zdziarski for helpful discussions. L.N.\ is grateful to Katelyn Breivik for advice on the \textsc{cosmic} code. L.N.\ and S.R.\ were partially supported by a BRICS STI grant and by a NITheCS grant from the National Research Foundation, South Africa. 
J.D.F.\ was supported by NASA through contract S-15633Y and the Office of Naval Research.
\end{acknowledgements}

\bibliographystyle{aa} 
\bibliography{UVXray_ref}

\begin{thebibliography}{65}
\expandafter\ifx\csname natexlab\endcsname\relax\def\natexlab#1{#1}\fi

\bibitem[{Bachetti {et~al.}(2014)Bachetti, Harrison, Walton, Grefenstette,
  Chakrabarty, F{\"u}rst, Barret, Beloborodov, Boggs, Christensen,
  {et~al.}}]{bachetti2014ultraluminous}
Bachetti, M., Harrison, F., Walton, D.~J., {et~al.} 2014, Nature, 514, 202

\bibitem[{{Bachetti} {et~al.}(2022){Bachetti}, {Heida}, {Maccarone},
  {Huppenkothen}, {Israel}, {Barret}, {Brightman}, {Brumback}, {Earnshaw},
  {Forster}, {F{\"u}rst}, {Grefenstette}, {Harrison}, {Jaodand}, {Madsen},
  {Middleton}, {Pike}, {Pilia}, {Poutanen}, {Stern}, {Tomsick}, {Walton},
  {Webb}, \& {Wilms}}]{2022ApJ...937..125B}
{Bachetti}, M., {Heida}, M., {Maccarone}, T., {et~al.} 2022, \apj, 937, 125

\bibitem[{{Begelman}(2001)}]{begelman01}
{Begelman}, M.~C. 2001, \apj, 551, 897

\bibitem[{{Belczynski} {et~al.}(2008){Belczynski}, {Kalogera}, {Rasio}, {Taam},
  {Zezas}, {Bulik}, {Maccarone}, \& {Ivanova}}]{belczynski2008compact}
{Belczynski}, K., {Kalogera}, V., {Rasio}, F.~A., {et~al.} 2008, \apjs, 174,
  223

\bibitem[{{Belczynski} {et~al.}(2007){Belczynski}, {Taam}, {Kalogera}, {Rasio},
  \& {Bulik}}]{belczynski2007rarity}
{Belczynski}, K., {Taam}, R.~E., {Kalogera}, V., {Rasio}, F.~A., \& {Bulik}, T.
  2007, \apj, 662, 504

\bibitem[{{Breivik} {et~al.}(2020){Breivik}, {Coughlin}, {Zevin}, {Rodriguez},
  {Kremer}, {Ye}, {Andrews}, {Kurkowski}, {Digman}, {Larson}, \&
  {Rasio}}]{breivik2020cosmic}
{Breivik}, K., {Coughlin}, S., {Zevin}, M., {et~al.} 2020, \apj, 898, 71

\bibitem[{{Brown} {et~al.}(2000){Brown}, {Lee}, {Wijers}, {Lee}, {Israelian},
  \& {Bethe}}]{brown2000theory}
{Brown}, G.~E., {Lee}, C.~H., {Wijers}, R.~A.~M.~J., {et~al.} 2000, \na, 5, 191

\bibitem[{{Colbert} \& {Mushotzky}(1999)}]{colbert1999nature}
{Colbert}, E. J.~M. \& {Mushotzky}, R.~F. 1999, \apj, 519, 89

\bibitem[{{Fabbiano}(1989)}]{fabbiano89}
{Fabbiano}, G. 1989, \araa, 27, 87

\bibitem[{{Finke} \& {B{\"o}ttcher}(2007)}]{finke07}
{Finke}, J.~D. \& {B{\"o}ttcher}, M. 2007, \apj, 667, 395

\bibitem[{{Finke} \& {Razzaque}(2017)}]{finke17}
{Finke}, J.~D. \& {Razzaque}, S. 2017, \mnras, 472, 3683

\bibitem[{{Fragos} {et~al.}(2023){Fragos}, {Andrews}, {Bavera}, {Berry},
  {Coughlin}, {Dotter}, {Giri}, {Kalogera}, {Katsaggelos}, {Kovlakas},
  {Lalvani}, {Misra}, {Srivastava}, {Qin}, {Rocha}, {Rom{\'a}n-Garza}, {Serra},
  {Stahle}, {Sun}, {Teng}, {Trajcevski}, {Tran}, {Xing}, {Zapartas}, \&
  {Zevin}}]{Fragos_2023}
{Fragos}, T., {Andrews}, J.~J., {Bavera}, S.~S., {et~al.} 2023, \apjs, 264, 45

\bibitem[{{Fryer} {et~al.}(2012){Fryer}, {Belczynski}, {Wiktorowicz},
  {Dominik}, {Kalogera}, \& {Holz}}]{fryer2012compact}
{Fryer}, C.~L., {Belczynski}, K., {Wiktorowicz}, G., {et~al.} 2012, \apj, 749,
  91

\bibitem[{{F{\"u}rst} {et~al.}(2016){F{\"u}rst}, {Walton}, {Harrison}, {Stern},
  {Barret}, {Brightman}, {Fabian}, {Grefenstette}, {Madsen}, {Middleton},
  {Miller}, {Pottschmidt}, {Ptak}, {Rana}, \& {Webb}}]{furst2016discovery}
{F{\"u}rst}, F., {Walton}, D.~J., {Harrison}, F.~A., {et~al.} 2016, \apjl, 831,
  L14

\bibitem[{{Gallegos-Garcia} {et~al.}(2021){Gallegos-Garcia}, {Berry},
  {Marchant}, \& {Kalogera}}]{Gallegos-Garcia_2021}
{Gallegos-Garcia}, M., {Berry}, C. P.~L., {Marchant}, P., \& {Kalogera}, V.
  2021, \apj, 922, 110

\bibitem[{{Georganopoulos} {et~al.}(2002){Georganopoulos}, {Aharonian}, \&
  {Kirk}}]{georganopoulos2002external}
{Georganopoulos}, M., {Aharonian}, F.~A., \& {Kirk}, J.~G. 2002, \aap, 388, L25

\bibitem[{{Gierli{\'n}ski} {et~al.}(2008){Gierli{\'n}ski}, {Done}, \&
  {Page}}]{2008MNRAS.388..753G}
{Gierli{\'n}ski}, M., {Done}, C., \& {Page}, K. 2008, \mnras, 388, 753

\bibitem[{{Gierli{\'n}ski} {et~al.}(2009){Gierli{\'n}ski}, {Done}, \&
  {Page}}]{2009MNRAS.392.1106G}
{Gierli{\'n}ski}, M., {Done}, C., \& {Page}, K. 2009, \mnras, 392, 1106

\bibitem[{{Gladstone} {et~al.}(2013){Gladstone}, {Copperwheat}, {Heinke},
  {Roberts}, {Cartwright}, {Levan}, \& {Goad}}]{gladstone2013optical}
{Gladstone}, J.~C., {Copperwheat}, C., {Heinke}, C.~O., {et~al.} 2013, \apjs,
  206, 14

\bibitem[{{Heger} {et~al.}(2003){Heger}, {Fryer}, {Woosley}, {Langer}, \&
  {Hartmann}}]{heger2003massive}
{Heger}, A., {Fryer}, C.~L., {Woosley}, S.~E., {Langer}, N., \& {Hartmann},
  D.~H. 2003, \apj, 591, 288

\bibitem[{{Heida} {et~al.}(2019){Heida}, {Lau}, {Davies}, {Brightman},
  {F{\"u}rst}, {Grefenstette}, {Kennea}, {Tramper}, {Walton}, \&
  {Harrison}}]{heida2019discovery}
{Heida}, M., {Lau}, R.~M., {Davies}, B., {et~al.} 2019, \apjl, 883, L34

\bibitem[{{Heida} {et~al.}(2015){Heida}, {Torres}, {Jonker}, {Servillat},
  {Repetto}, {Roberts}, {Walton}, {Moon}, \& {Harrison}}]{heida2015discovery}
{Heida}, M., {Torres}, M.~A.~P., {Jonker}, P.~G., {et~al.} 2015, \mnras, 453,
  3510

\bibitem[{{Inoue} {et~al.}(2016){Inoue}, {Tanaka}, \& {Isobe}}]{inoue16}
{Inoue}, Y., {Tanaka}, Y.~T., \& {Isobe}, N. 2016, \mnras, 461, 4329

\bibitem[{{Kaaret} {et~al.}(2017){Kaaret}, {Feng}, \&
  {Roberts}}]{kaaret2017ultraluminous}
{Kaaret}, P., {Feng}, H., \& {Roberts}, T.~P. 2017, \araa, 55, 303

\bibitem[{King {et~al.}(2023)King, Lasota, \& Middleton}]{KING2023101672}
King, A., Lasota, J.-P., \& Middleton, M. 2023, New Astronomy Reviews, 96,
  101672

\bibitem[{{King}(2009)}]{king2009masses}
{King}, A.~R. 2009, \mnras, 393, L41

\bibitem[{{King} {et~al.}(2001){King}, {Davies}, {Ward}, {Fabbiano}, \&
  {Elvis}}]{king2001ultraluminous}
{King}, A.~R., {Davies}, M.~B., {Ward}, M.~J., {Fabbiano}, G., \& {Elvis}, M.
  2001, \apjl, 552, L109

\bibitem[{{Kosec} {et~al.}(2018){Kosec}, {Pinto}, {Walton}, {Fabian},
  {Bachetti}, {Brightman}, {F{\"u}rst}, \&
  {Grefenstette}}]{10.1093/mnras/sty1626}
{Kosec}, P., {Pinto}, C., {Walton}, D.~J., {et~al.} 2018, \mnras, 479, 3978

\bibitem[{{Kroupa} {et~al.}(1993){Kroupa}, {Tout}, \&
  {Gilmore}}]{kroupa1993distribution}
{Kroupa}, P., {Tout}, C.~A., \& {Gilmore}, G. 1993, \mnras, 262, 545

\bibitem[{{Lasota} {et~al.}(2011){Lasota}, {Alexander}, {Dubus}, {Barret},
  {Farrell}, {Gehrels}, {Godet}, \& {Webb}}]{lasota2011origin}
{Lasota}, J.~P., {Alexander}, T., {Dubus}, G., {et~al.} 2011, \apj, 735, 89

\bibitem[{Lasota {et~al.}(2016)Lasota, Vieira, Sadowski, Narayan, \&
  Abramowicz}]{lasota2016slimming}
Lasota, J.-P., Vieira, R., Sadowski, A., Narayan, R., \& Abramowicz, M. 2016,
  \aap, 587, A13

\bibitem[{{Lehmer} {et~al.}(2021){Lehmer}, {Eufrasio}, {Basu-Zych}, {Doore},
  {Fragos}, {Garofali}, {Kovlakas}, {Williams}, {Zezas}, \&
  {Santana-Silva}}]{lehmer2021metallicity}
{Lehmer}, B.~D., {Eufrasio}, R.~T., {Basu-Zych}, A., {et~al.} 2021, \apj, 907,
  17

\bibitem[{{Liu} {et~al.}(2004){Liu}, {Bregman}, \& {Seitzer}}]{liu2004optical}
{Liu}, J.-F., {Bregman}, J.~N., \& {Seitzer}, P. 2004, \apj, 602, 249

\bibitem[{{Mapelli} {et~al.}(2011){Mapelli}, {Ripamonti}, {Zampieri}, \&
  {Colpi}}]{mapelli2011remnants}
{Mapelli}, M., {Ripamonti}, E., {Zampieri}, L., \& {Colpi}, M. 2011,
  Astronomische Nachrichten, 332, 414

\bibitem[{{Marchant} {et~al.}(2021){Marchant}, {Pappas}, {Gallegos-Garcia},
  {Berry}, {Taam}, {Kalogera}, \& {Podsiadlowski}}]{2021A&A...650A.107M}
{Marchant}, P., {Pappas}, K. M.~W., {Gallegos-Garcia}, M., {et~al.} 2021, \aap,
  650, A107

\bibitem[{McKinney {et~al.}(2014)McKinney, Tchekhovskoy, Sadowski, \&
  Narayan}]{mckinney2014three}
McKinney, J.~C., Tchekhovskoy, A., Sadowski, A., \& Narayan, R. 2014, \mnras,
  441, 3177

\bibitem[{Mezcua {et~al.}(2013)Mezcua, Roberts, Sutton, \&
  Lobanov}]{mezcua2013radio}
Mezcua, M., Roberts, T., Sutton, A., \& Lobanov, A. 2013, \mnras, 436, 3128

\bibitem[{Middleton {et~al.}(2019)Middleton, Brightman, Pintore, Bachetti,
  Fabian, Fuerst, \& Walton}]{middleton2019magnetic}
Middleton, M., Brightman, M., Pintore, F., {et~al.} 2019, \mnras, 486, 2

\bibitem[{Middleton {et~al.}(2015)Middleton, Heil, Pintore, Walton, \&
  Roberts}]{middleton2015spectral}
Middleton, M.~J., Heil, L., Pintore, F., Walton, D.~J., \& Roberts, T.~P. 2015,
  \mnras, 447, 3243

\bibitem[{Middleton {et~al.}(2013)Middleton, Miller-Jones, Markoff, Fender,
  Henze, Hurley-Walker, Scaife, Roberts, Walton, Carpenter,
  {et~al.}}]{middleton2013bright}
Middleton, M.~J., Miller-Jones, J.~C., Markoff, S., {et~al.} 2013, Nature, 493,
  187

\bibitem[{{Mitsuda} {et~al.}(1984){Mitsuda}, {Inoue}, {Koyama}, {Makishima},
  {Matsuoka}, {Ogawara}, {Shibazaki}, {Suzuki}, {Tanaka}, \&
  {Hirano}}]{1984PASJ...36..741M}
{Mitsuda}, K., {Inoue}, H., {Koyama}, K., {et~al.} 1984, \pasj, 36, 741

\bibitem[{{Mondal} {et~al.}(2020){Mondal}, {Belczy{\'n}ski}, {Wiktorowicz},
  {Lasota}, \& {King}}]{mondal20}
{Mondal}, S., {Belczy{\'n}ski}, K., {Wiktorowicz}, G., {Lasota}, J.-P., \&
  {King}, A.~R. 2020, \mnras, 491, 2747

\bibitem[{Motch {et~al.}(2014)Motch, Pakull, Soria, Gris{\'e}, \&
  Pietrzy{\'n}ski}]{motch2014mass}
Motch, C., Pakull, M., Soria, R., Gris{\'e}, F., \& Pietrzy{\'n}ski, G. 2014,
  Nature, 514, 198

\bibitem[{Murdin \& Webster(1971)}]{murdin1971optical}
Murdin, P. \& Webster, B.~L. 1971, Nature, 233, 110

\bibitem[{{Mushtukov} {et~al.}(2021){Mushtukov}, {Portegies Zwart},
  {Tsygankov}, {Nagirner}, \& {Poutanen}}]{2021MNRAS.501.2424M}
{Mushtukov}, A.~A., {Portegies Zwart}, S., {Tsygankov}, S.~S., {Nagirner},
  D.~I., \& {Poutanen}, J. 2021, \mnras, 501, 2424

\bibitem[{Patton \& Sukhbold(2020)}]{10.1093/mnras/staa3029}
Patton, R.~A. \& Sukhbold, T. 2020, \mnras, 499, 2803

\bibitem[{{Paxton} {et~al.}(2011){Paxton}, {Bildsten}, {Dotter}, {Herwig},
  {Lesaffre}, \& {Timmes}}]{Paxton_2010}
{Paxton}, B., {Bildsten}, L., {Dotter}, A., {et~al.} 2011, \apjs, 192, 3

\bibitem[{Pinto {et~al.}(2020)Pinto, Walton, Kara, Parker, Soria, Kosec,
  Middleton, Alston, Fabian, Guainazzi, {et~al.}}]{pinto2020xmm}
Pinto, C., Walton, D., Kara, E., {et~al.} 2020, \mnras, 492, 4646

\bibitem[{{Pinto} \& {Walton}(2023)}]{pinto2023ultra}
{Pinto}, C. \& {Walton}, D.~J. 2023, arXiv e-prints, arXiv:2302.00006

\bibitem[{Poutanen {et~al.}(2007)Poutanen, Lipunova, Fabrika, Butkevich, \&
  Abolmasov}]{poutanen2007supercritically}
Poutanen, J., Lipunova, G., Fabrika, S., Butkevich, A.~G., \& Abolmasov, P.
  2007, \mnras, 377, 1187

\bibitem[{Roberts {et~al.}(2008)Roberts, Levan, \& Goad}]{roberts2008new}
Roberts, T., Levan, A., \& Goad, M. 2008, \mnras, 387, 73

\bibitem[{S{\k{a}}dowski \& Narayan(2015)}]{skadowski2015powerful}
S{\k{a}}dowski, A. \& Narayan, R. 2015, \mnras, 453, 3213

\bibitem[{{Salvaggio} {et~al.}(2023){Salvaggio}, {Wolter}, {Belfiore}, \&
  {Colpi}}]{salvaggio23}
{Salvaggio}, C., {Wolter}, A., {Belfiore}, A., \& {Colpi}, M. 2023, \mnras,
  522, 1377

\bibitem[{Sana {et~al.}(2013)Sana, de~Koter, De~Mink, Dunstall, Evans,
  H{\'e}nault-Brunet, Apell{\'a}niz, Ram{\'\i}rez-Agudelo, Taylor, Walborn,
  {et~al.}}]{sana2013vlt}
Sana, H., de~Koter, A., De~Mink, S., {et~al.} 2013, \aap, 550, A107

\bibitem[{{Shakura} \& {Sunyaev}(1973)}]{shakura1973black}
{Shakura}, N.~I. \& {Sunyaev}, R.~A. 1973, \aap, 24, 337

\bibitem[{{Sonbas} {et~al.}(2019){Sonbas}, {Dhuga}, \&
  {G{\"o}{\u{g}}{\"u}{\c{s}}}}]{sonbas2019evidence}
{Sonbas}, E., {Dhuga}, K.~S., \& {G{\"o}{\u{g}}{\"u}{\c{s}}}, E. 2019, \apjl,
  873, L12

\bibitem[{{Tananbaum} {et~al.}(1979){Tananbaum}, {Avni}, {Branduardi}, {Elvis},
  {Fabbiano}, {Feigelson}, {Giacconi}, {Henry}, {Pye}, {Soltan}, \&
  {Zamorani}}]{tananbaum1979x}
{Tananbaum}, H., {Avni}, Y., {Branduardi}, G., {et~al.} 1979, \apjl, 234, L9

\bibitem[{Tao {et~al.}(2012)Tao, Feng, Kaaret, Grisé, \& Jin}]{Tao_2012}
Tao, L., Feng, H., Kaaret, P., Grisé, F., \& Jin, J. 2012, \apj, 758, 85

\bibitem[{Vagnozzi {et~al.}(2017)Vagnozzi, Freese, \&
  Zurbuchen}]{vagnozzi2017solar}
Vagnozzi, S., Freese, K., \& Zurbuchen, T.~H. 2017, \apj, 839, 55

\bibitem[{Van Den~Eijnden {et~al.}(2019)Van Den~Eijnden, Degenaar, Schulz,
  Nowak, Wijnands, Russell, Hern{\'a}ndez~Santisteban, Bahramian, Maccarone,
  Kennea, {et~al.}}]{van2019chandra}
Van Den~Eijnden, J., Degenaar, N., Schulz, N., {et~al.} 2019, \mnras, 487, 4355

\bibitem[{Vasilopoulos {et~al.}(2020)Vasilopoulos, Ray, Gendreau, Jenke,
  Jaisawal, Wilson-Hodge, Strohmayer, Altamirano, Iwakiri, Wolff,
  {et~al.}}]{vasilopoulos20202019}
Vasilopoulos, G., Ray, P., Gendreau, K., {et~al.} 2020, \mnras, 494, 5350

\bibitem[{{Vinokurov} {et~al.}(2013){Vinokurov}, {Fabrika}, \&
  {Atapin}}]{vinokurov2013ultra}
{Vinokurov}, A., {Fabrika}, S., \& {Atapin}, K. 2013, Astrophysical Bulletin,
  68, 139

\bibitem[{Walton {et~al.}(2022)Walton, Mackenzie, Gully, Patel, Roberts,
  Earnshaw, \& Mateos}]{walton2022multimission}
Walton, D., Mackenzie, A., Gully, H., {et~al.} 2022, \mnras, 509, 1587

\bibitem[{Walton {et~al.}(2011)Walton, Roberts, Mateos, \&
  Heard}]{walton20112xmm}
Walton, D., Roberts, T., Mateos, S., \& Heard, V. 2011, \mnras, 416, 1844

\bibitem[{Wiktorowicz {et~al.}(2017)Wiktorowicz, Sobolewska, Lasota, \&
  Belczynski}]{wiktorowicz2017origin}
Wiktorowicz, G., Sobolewska, M., Lasota, J.-P., \& Belczynski, K. 2017, \apj,
  846, 17

\end{thebibliography}

\begin{appendix}

\section*{Appendix A}

\renewcommand{\thefigure}{A\arabic{figure}}
{Fig.~\ref{fig:SED_mdot} shows the SED for a $15~\text{M}_{\odot}$ BH and a $1.4~\text{M}_{\odot}$ NS with varying accretion rate ratio ($\dot{m}$). This illustrates that the X-ray emission at 2 keV is not affected by changing $\dot{m}$ for a given mass ($M$), indicating that the X-ray flux is dominated by emission from the inner disk in the SCAD model.}

 Figure~\ref{fig:sed_02c} shows an SED from the SCAD model with a higher wind velocity (60,000 km/s or $0.2c$), compared to what we assume in most of this paper (1000 km/s).  Wind velocity $v_w$ affects the photosphere radius $r_{\rm ph}$, which is the outer radius of the disk; see the text after Equation (\ref{Vsp}).  A higher $v_w$ leads to a smaller $r_{\rm ph}$, truncating the disk at a lower distance.  This affects the UV portion of the spectrum.  The X-rays from the inner part of the disk are unaffected, as seen in Fig.~\ref{fig:sed_02c}.

Figure~\ref{fig:SEDfout} shows the effect of the funnel opening angle $\theta_f$ on the SED.   Changes in $f_{\rm out}$ affect only the UV component, but not the X-ray component.
 On the other hand, changes in $\theta_f$ affect the ``shelf'' of the X-ray component but not the UV component.

Figure~\ref{fig:alphafout} shows $\alpha_{\rm ox} - L_{\nu,\rm UV}$ relation for different values of the parameter $f_{\rm out}$.  The effect is minimal, since the changes in $f_{\rm out}$ do not affect the X-ray component of the spectrum. \\

Figure~\ref{fig:dist_beam} shows the distribution of the beaming factor in our ULX populations of different metallicity. Changes in metallicity do not have a strong effect on the distribution of beaming factors. Figure~\ref{fig:obse_dist_beam} shows the distribution of the beaming factor in our ULX populations of different metallicity for the observable population.

\begin{figure}[H]
    \centering
    \includegraphics[width=1\linewidth]{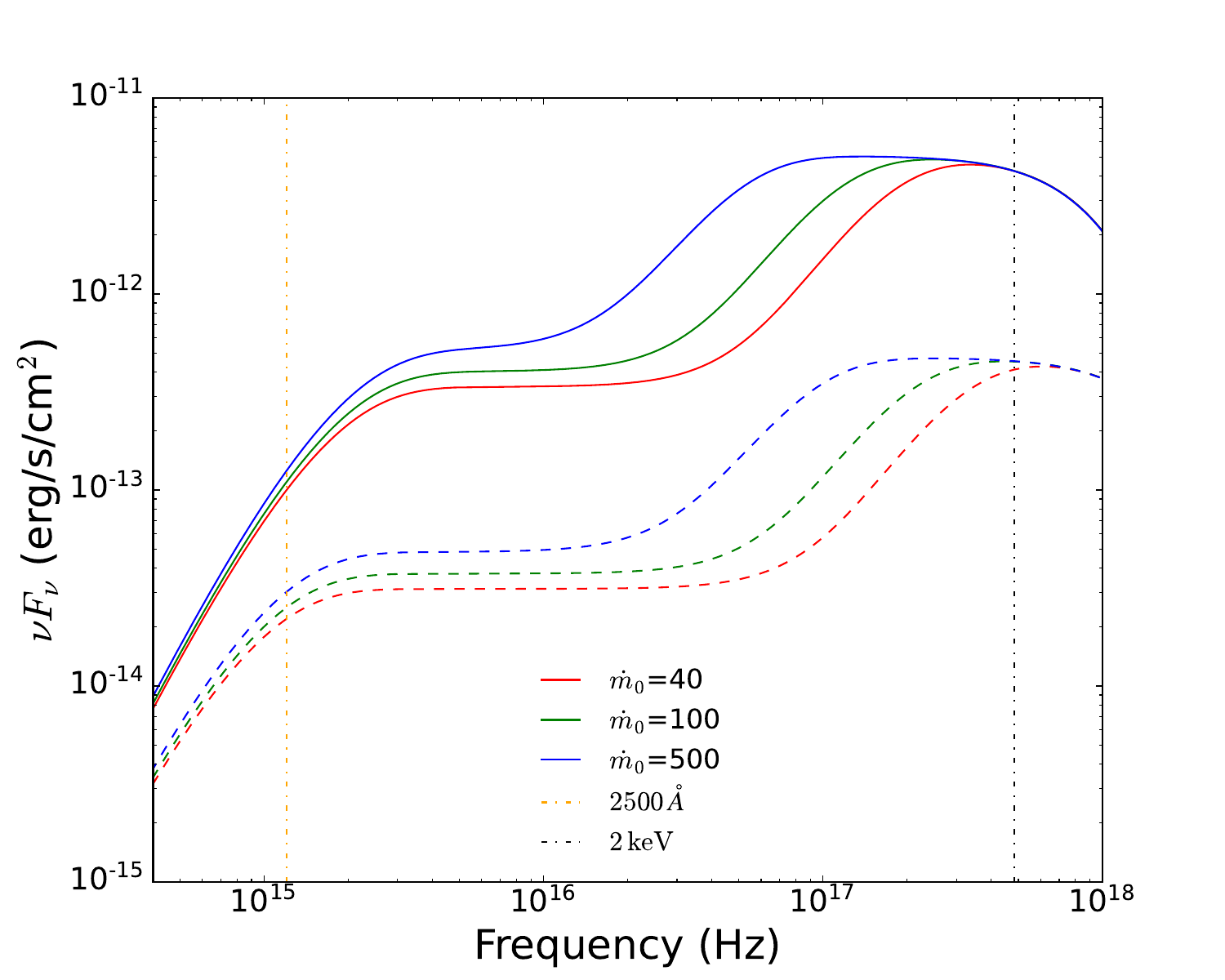}
        \caption{The SEDs of the accretion disk for a BH with $M= 15~\text{M}_{\odot}$(solid lines) and a NS with  $M=1.4~\text{M}_{\odot}$ (dashed lines), and for diﬀerent accretion rates $\dot{m}=40$ (red), 100 (green), and 500 (blue). In all cases $f_{\rm out}=0.03$. The vertical dot-dashed lines show the frequency at $2500\text{\AA}$ and $2$ keV. }
    \label{fig:SED_mdot}
\end{figure}

\begin{figure}[H]
    \centering
    \includegraphics[scale=0.38]{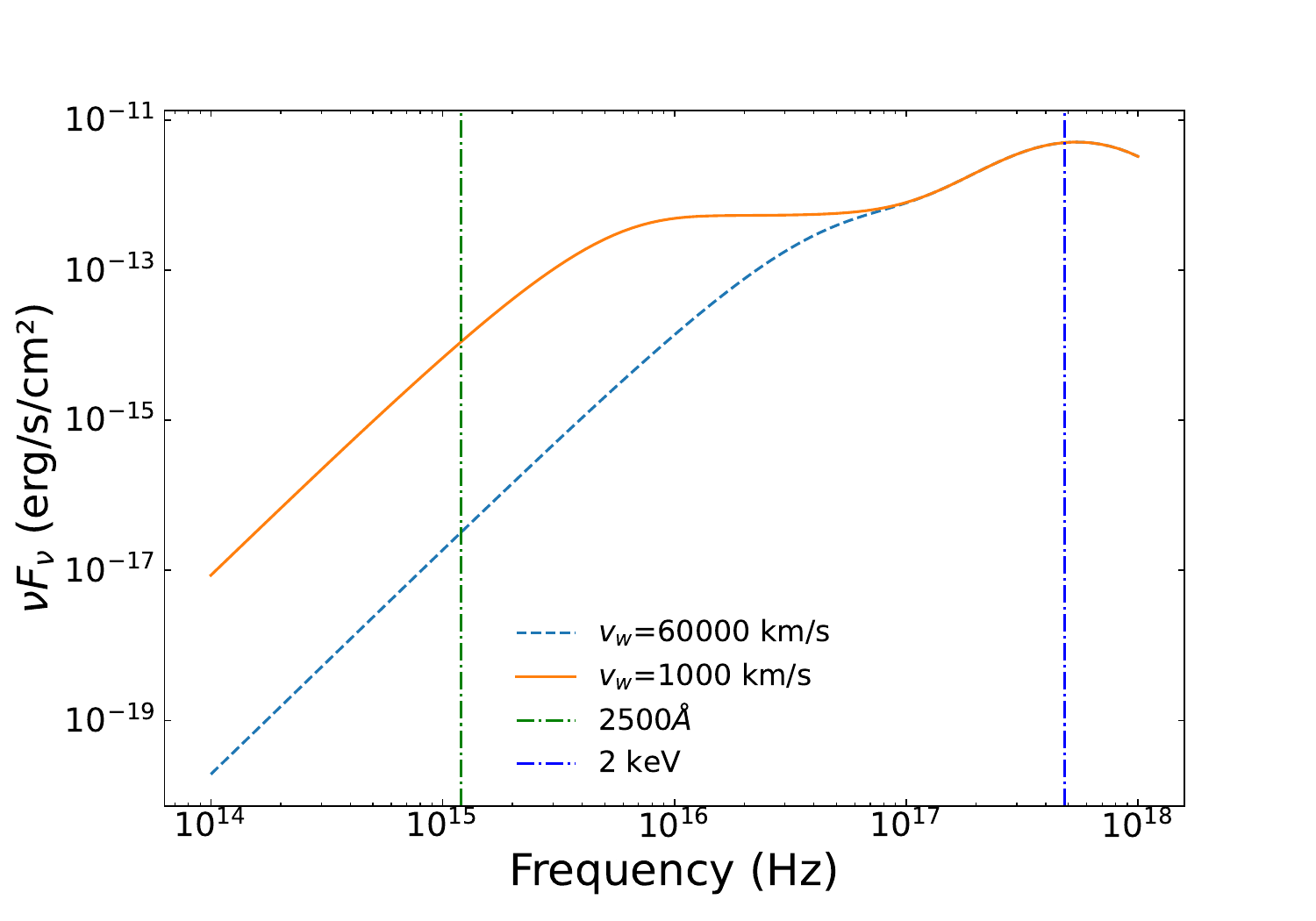}
    \caption{SED with wind velocities 0.2c (60 000 km/s) dashed line and c/300 (1000 km/s) solid line for a BH-ULX system with $M=10$ M$_{\odot}$, $\dot{m}=14$, $\theta_f=45^{\circ}$, and $f_{\rm out}=0.03$, indicating a reduction of the UV shoulder while maintaining the same flux in the X-ray band.}
    \label{fig:sed_02c}
\end{figure}

\begin{figure}[H]
    \centering
    \includegraphics[scale=0.38]{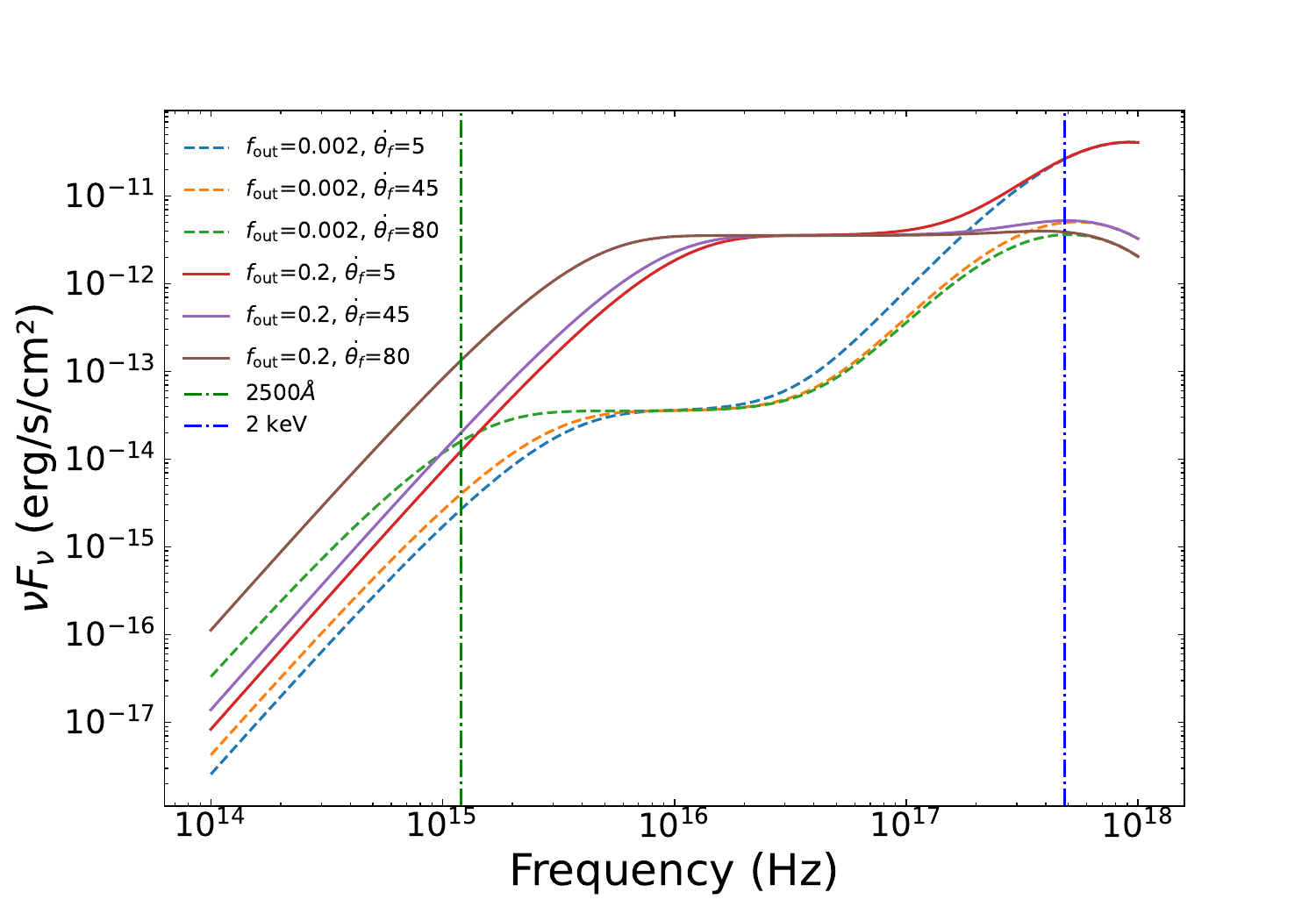}
    \caption{ The SEDs of $M_{\rm BH}=10~{\rm M}_{\odot}$ and $\dot{m}=14$ BH-ULX system. The plot shows how varying the thermalisation fraction $f_{\rm out}$ (the dashed line is for $f_{\rm out}=0.002$ and solid line is for $f_{\rm out}=0.2$) and funnel opening angle $\theta_f$ affects the emitted flux across different frequencies. }
    \label{fig:SEDfout}
\end{figure}

\begin{figure}
    \centering
    \begin{subfigure}{0.5\textwidth}
        \centering
        \includegraphics[scale=0.3]{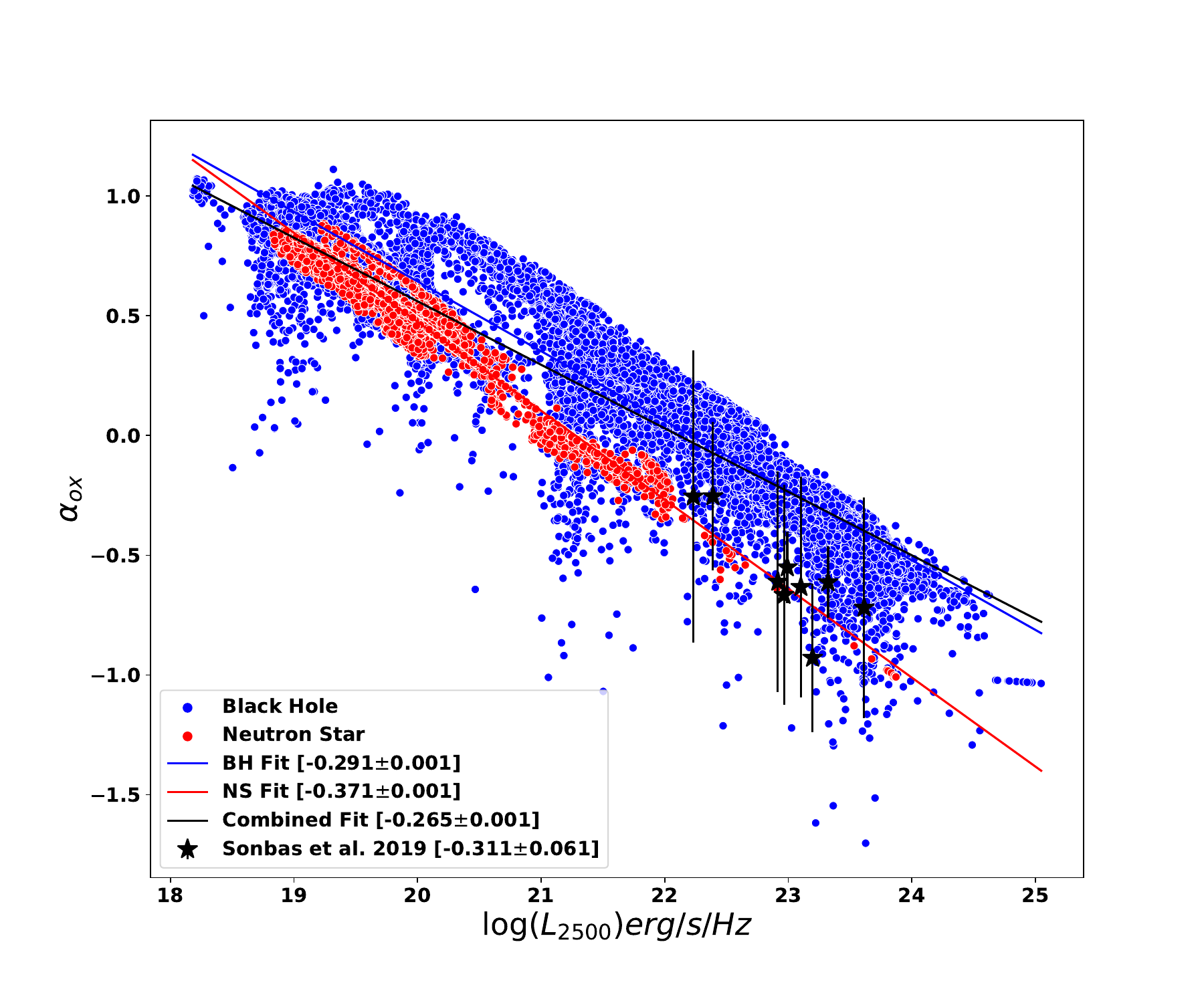}
        \vspace{-0.6cm}
        \label{fig:subfig-a}
    \end{subfigure}
    \hfill
    \begin{subfigure}{0.5\textwidth}
        \centering
        \vspace{-0.6cm}
        \includegraphics[scale=0.3]{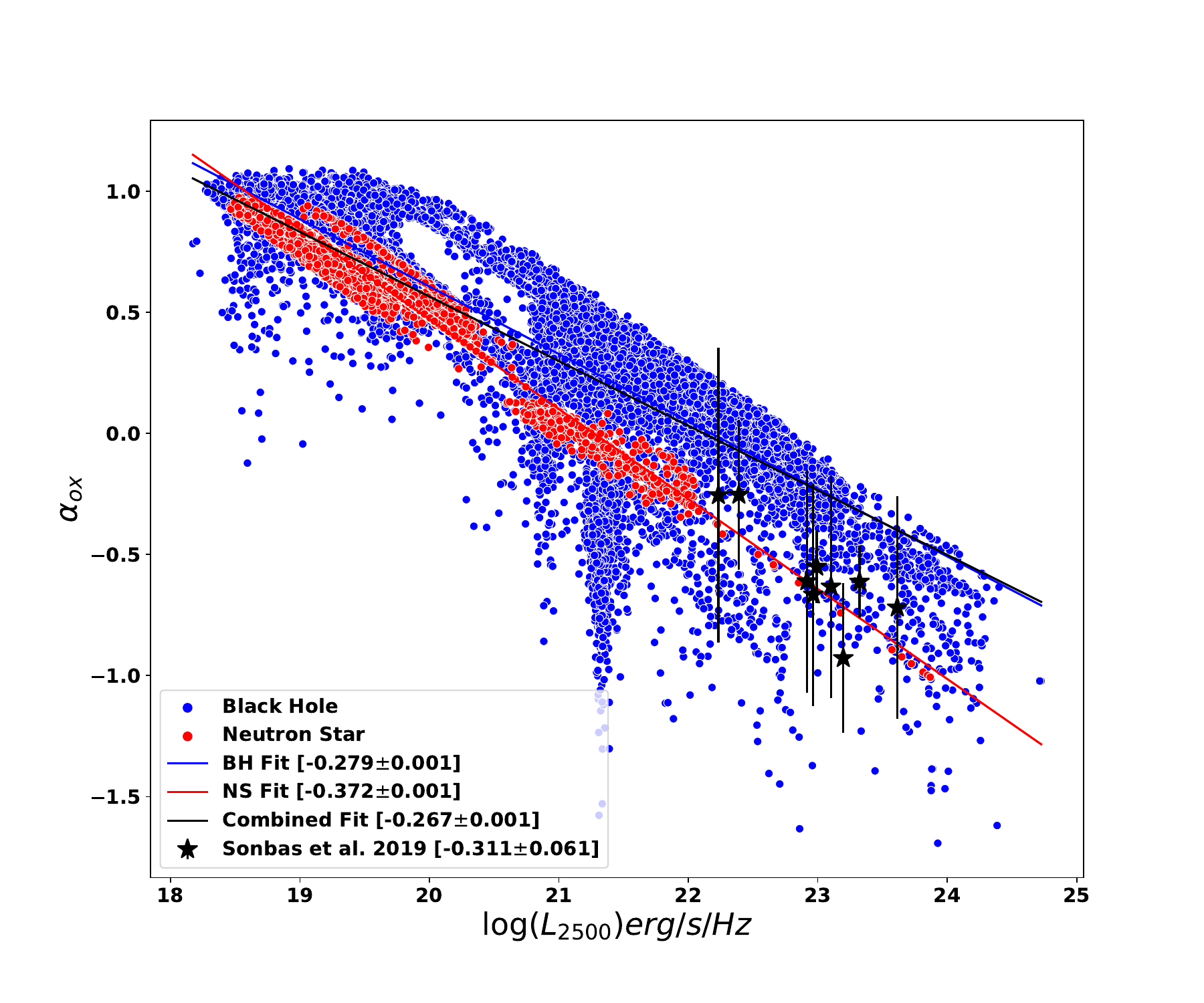}
         \vspace{-0.5cm}
        \label{fig:subfig-b}
    \end{subfigure}
    \hfill
    \begin{subfigure}{0.5\textwidth}
        \centering
        \vspace{-0.6cm}
        \includegraphics[scale=0.3]{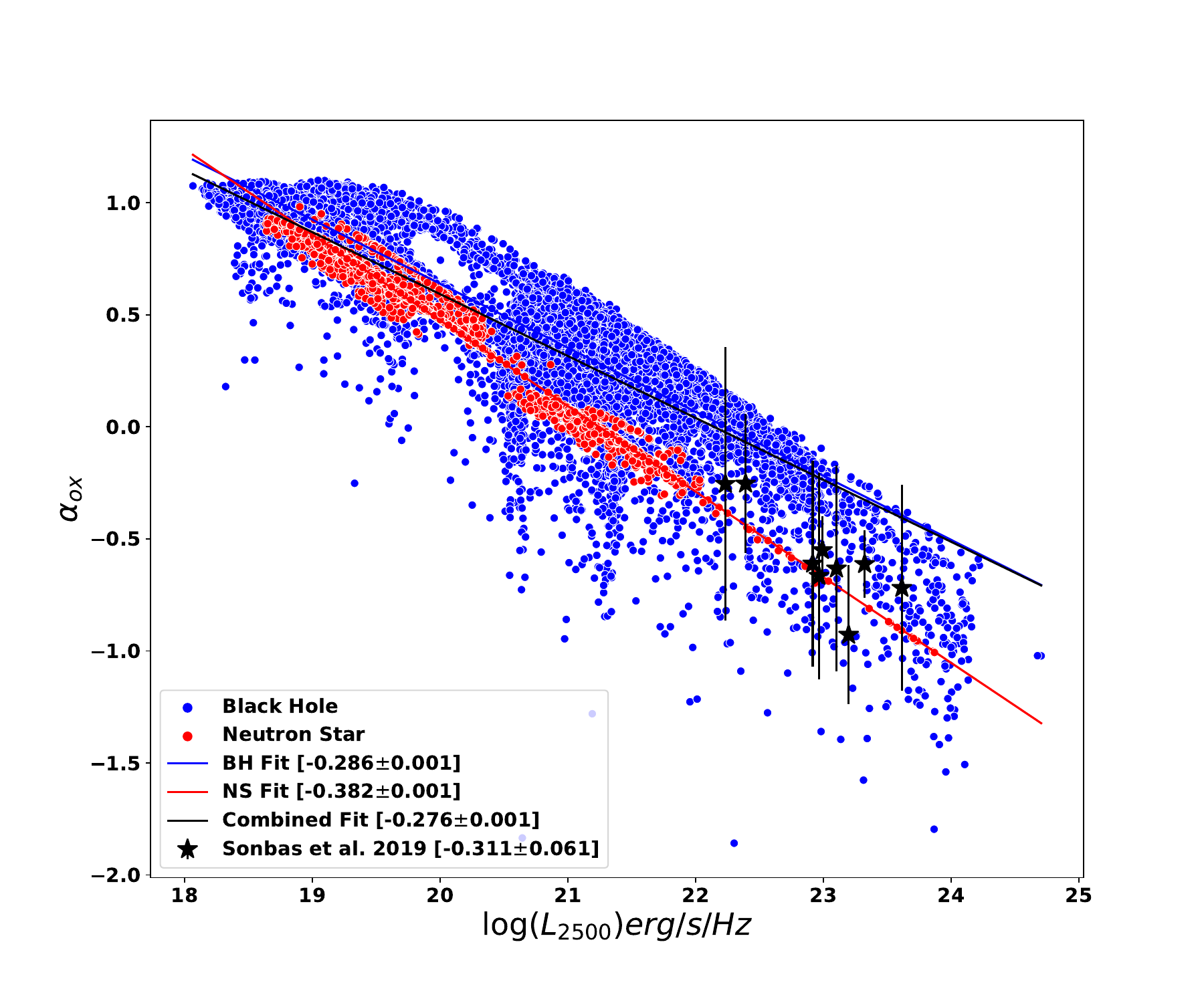}
        \vspace{-0.6cm}
        \label{fig:subfig-c}
    \end{subfigure}
    \caption{{$\alpha_{\rm ox}$ index as a function of $L_{\nu,\rm UV}$ (2500 Å) luminosity for our simulated ULX population (BH or NS primaries) with disc wind velocity $v_{\rm w}=60000$ km/s and $\theta_f=45^{\circ}$. The best-fit to $\alpha_{\rm ox} - L_{\nu,\rm UV}$ simulated data in our population (black solid line) and observed ULXs data in \citet{sonbas2019evidence} (black points) are also shown. The top, middle and bottom panel correspond to metallicities of $0.25Z_\odot$, $0.2Z_\odot$, and $0.4Z_\odot$, respectively. }}
  \label{fig:alpha_02c}
\end{figure}

\begin{figure}
    \centering
    \begin{subfigure}{0.5\textwidth}
        \centering
        \includegraphics[scale=0.3]{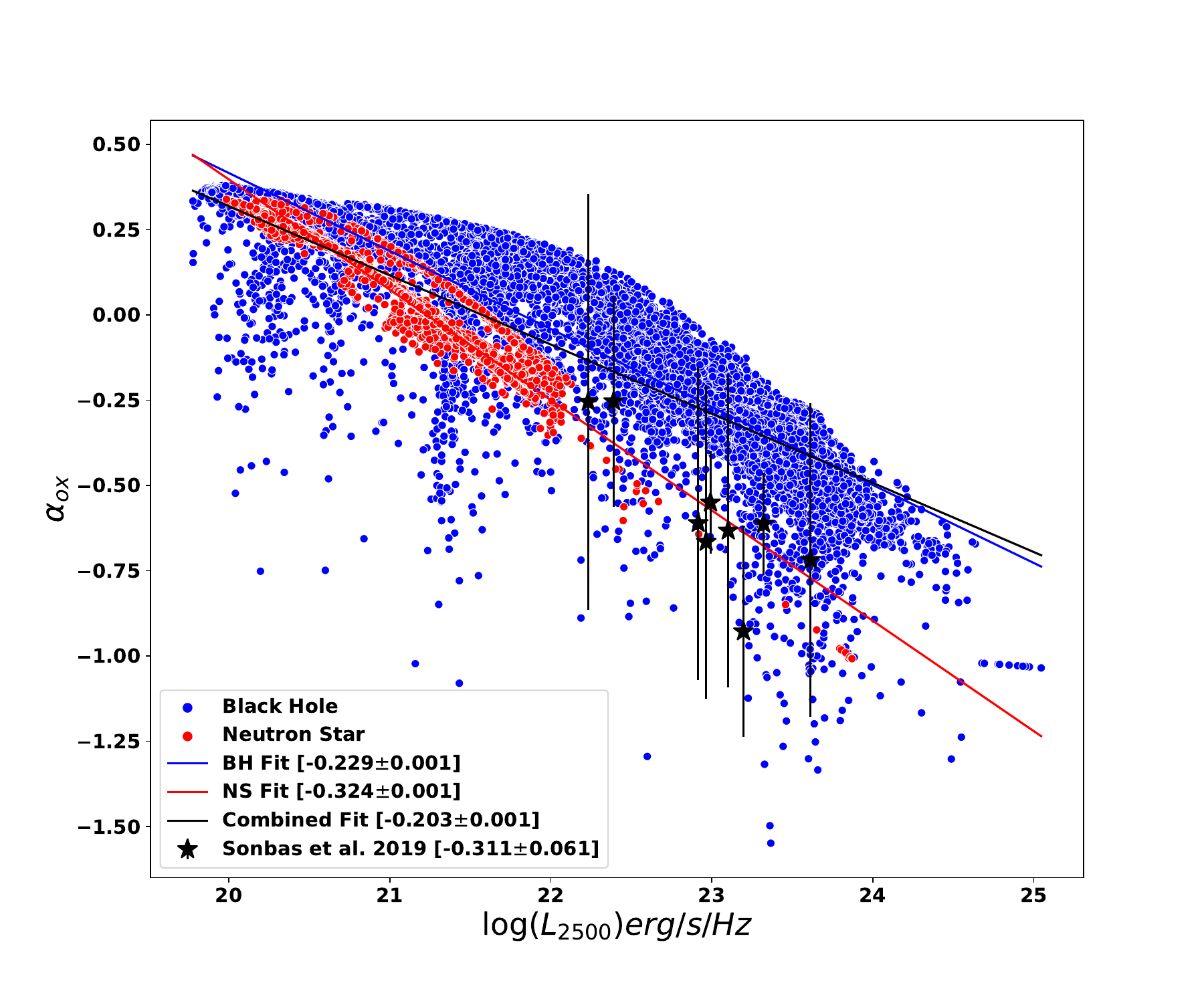}
        \vspace{-0.6cm}
        \label{fig:subfig-a}
    \end{subfigure}
    \hfill
    \begin{subfigure}{0.5\textwidth}
        \centering
        \vspace{-0.6cm}
        \includegraphics[scale=0.3]{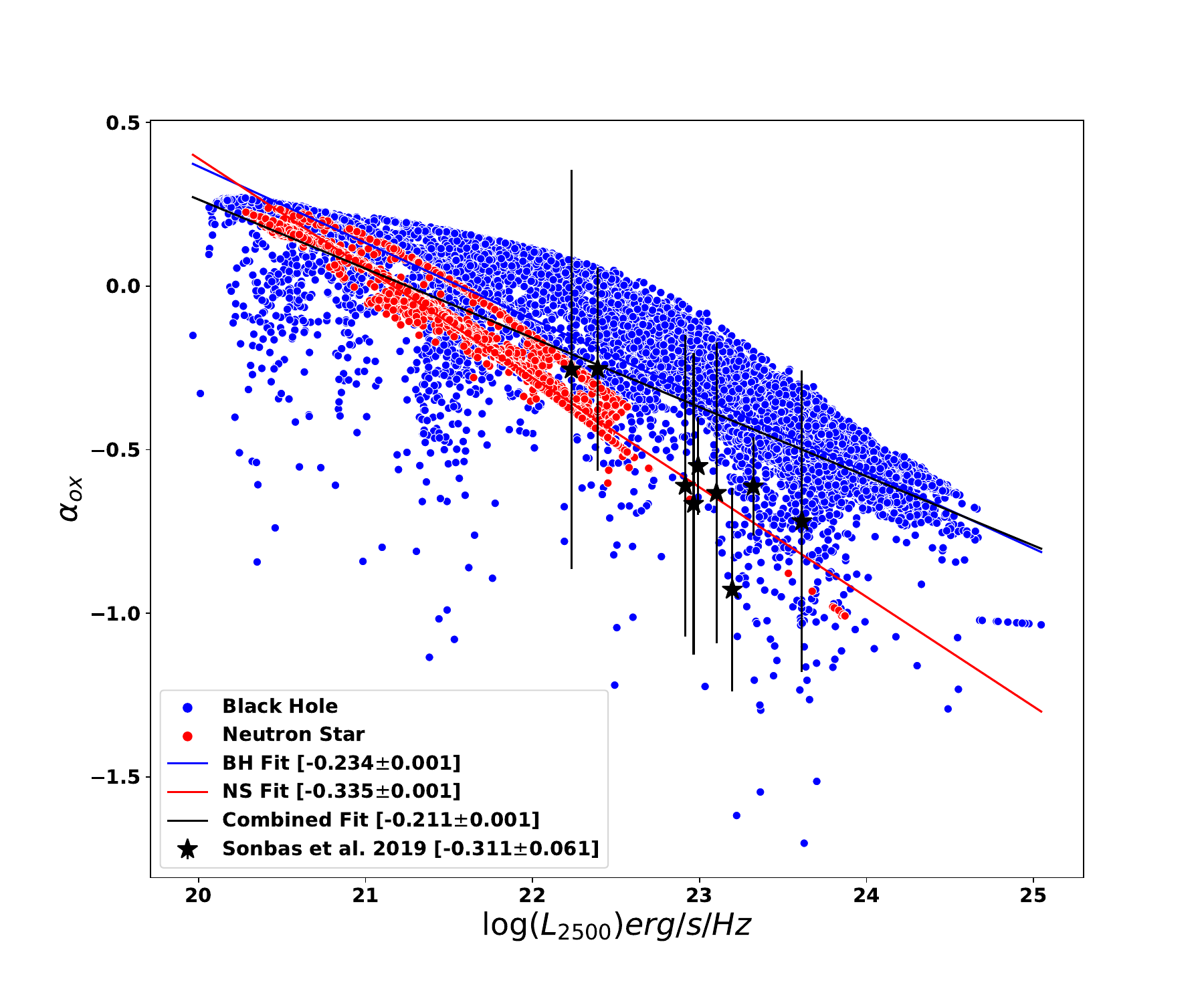}
         \vspace{-0.5cm}
        \label{fig:subfig-b}
    \end{subfigure}
    \hfill
    \begin{subfigure}{0.5\textwidth}
        \centering
        \vspace{-0.6cm}
        \includegraphics[scale=0.3]{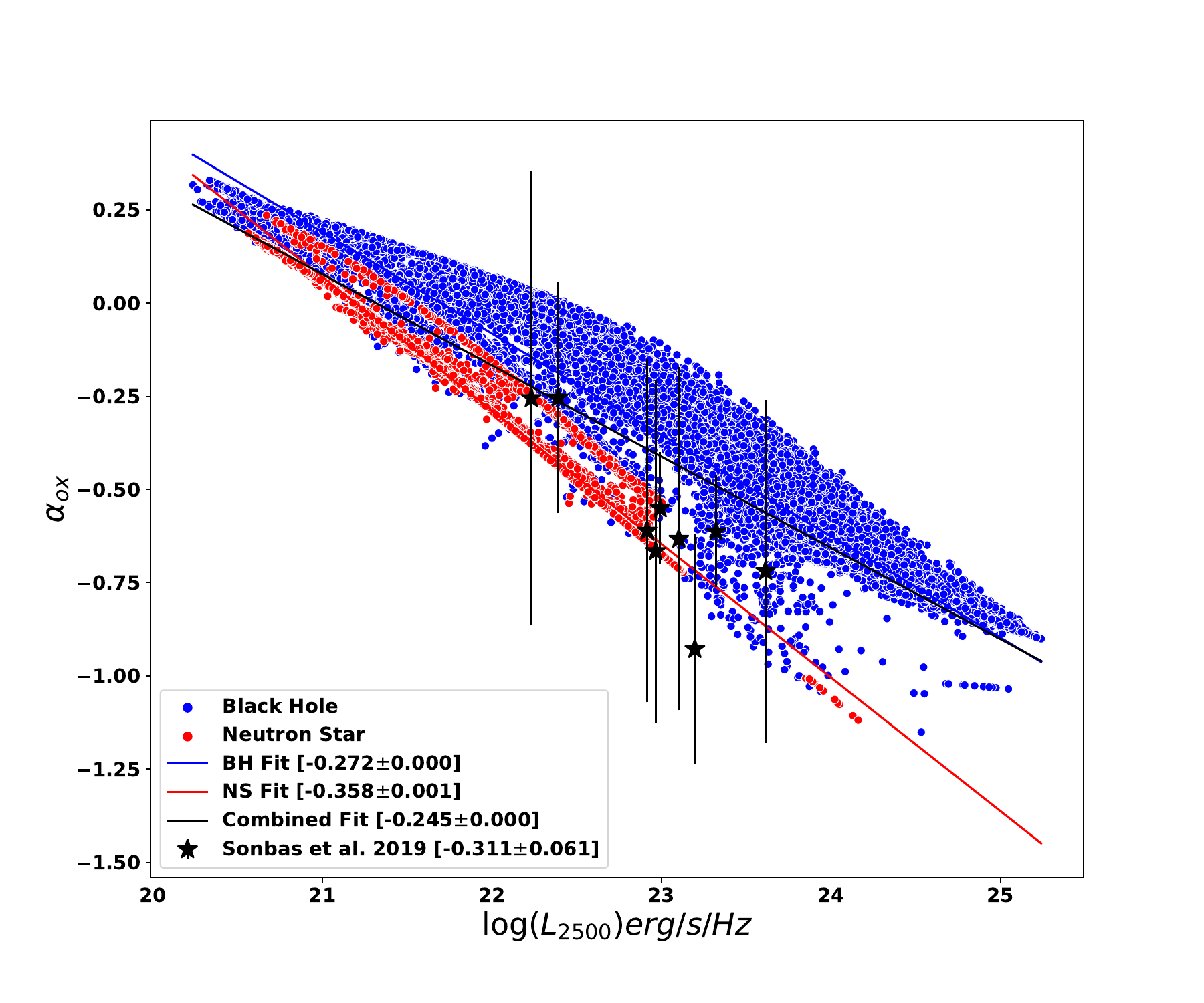}
        \vspace{-0.6cm}
        \label{fig:subfig-c}
    \end{subfigure}
    \caption{$\alpha_{\rm ox}$ index as a function of $L_{\nu,\rm UV}$ (2500 Å) luminosity for our simulated ULX population (BH or NS primaries) at $0.025Z_\odot$ without beaming included in the calculation. The plots show results for different values of $f_{\rm out}$. The top, middle, and bottom panels correspond to $f_{\rm out}$ values of 0.002, 0.03, and 0.2, respectively. The best fit to $\alpha_{\rm ox} - L_{\nu,\rm UV}$ simulated data in our population (black solid line) and observed ULX data from \citet{sonbas2019evidence} (black points) are also shown. }
    \label{fig:alphafout}
\end{figure}

\begin{figure*}
    \centering
    \includegraphics[width=1.0\linewidth]{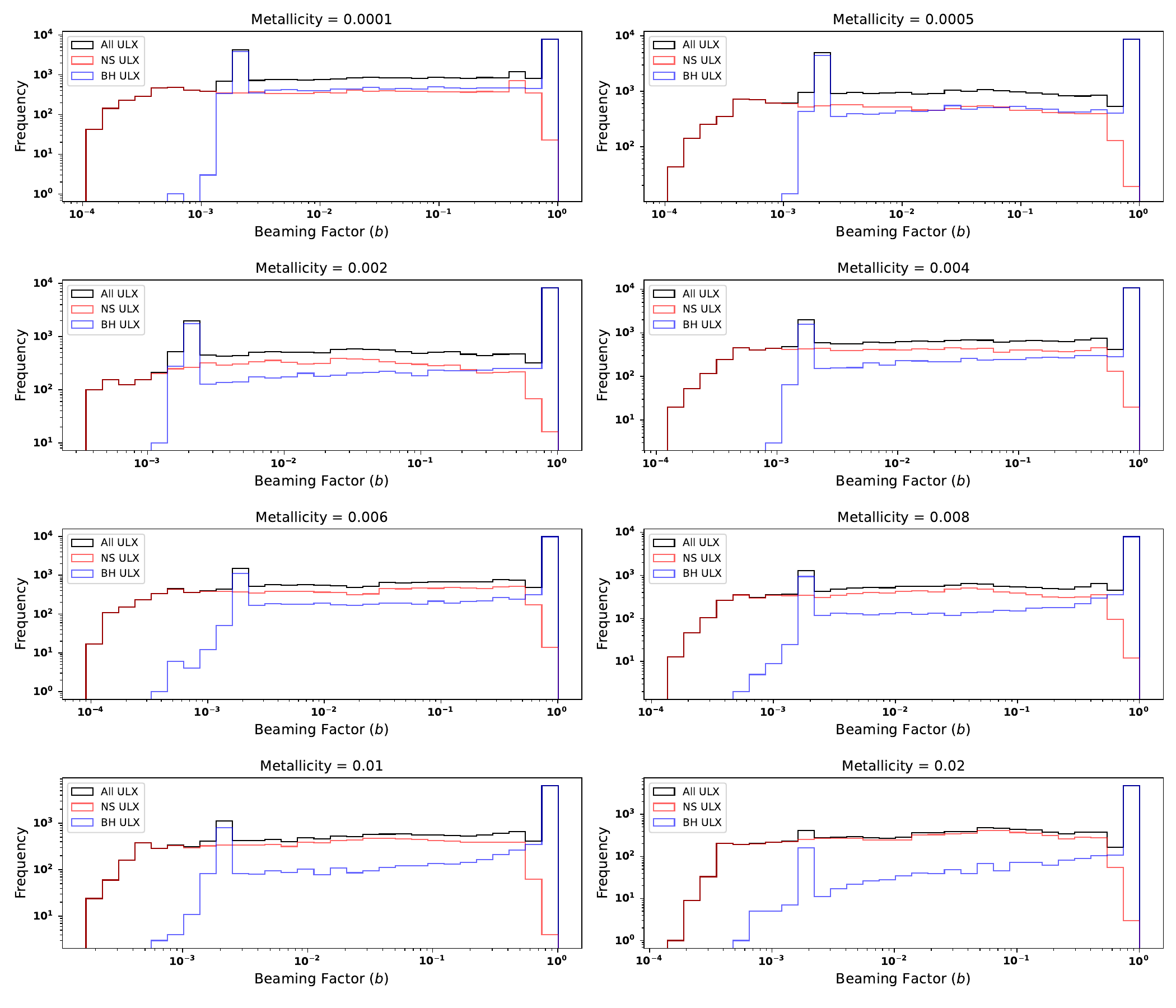}
    \caption{ Histograms of beaming factors across different metallicity environments.}
    \label{fig:dist_beam}
\end{figure*}

\begin{figure*}
    \centering
    \includegraphics[width=1.0\linewidth]{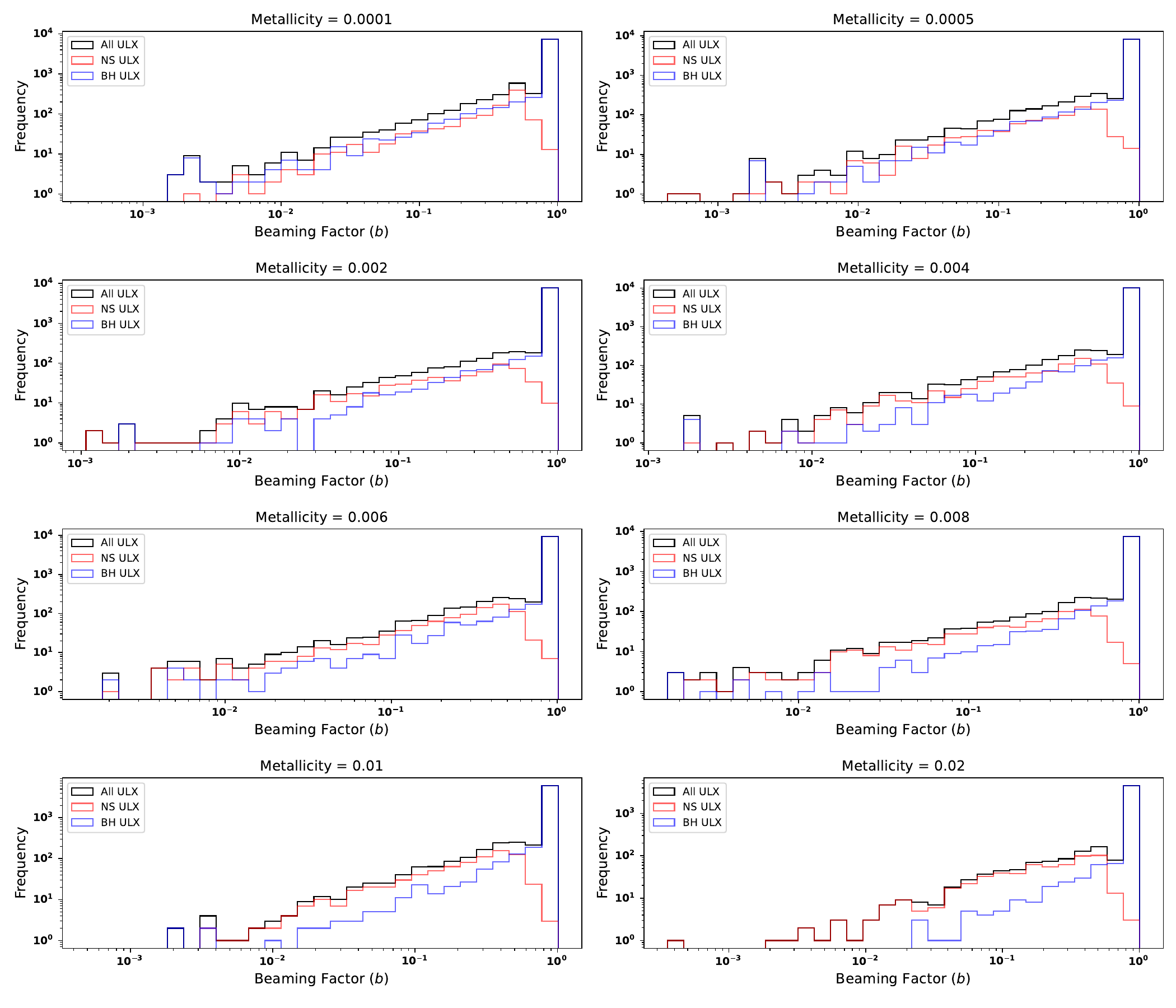}
    \caption{ Histograms of beaming factors for observable ULX population across different metallicity environments.}
    \label{fig:obse_dist_beam}
\end{figure*}

\end{appendix}

%
%

\end{document}